\documentclass[aps,pra,twocolumn,groupedaddress]{revtex4-2}

\usepackage{amsmath, amssymb}
\usepackage{bm}
\usepackage{lipsum}
\usepackage{mathrsfs}
\usepackage{xcolor}
\usepackage{graphicx}
\usepackage[separate-uncertainty]{siunitx}
\usepackage{physics2}
\usepackage{derivative}
\usepackage{braket}
\usepackage{mleftright}
\usepackage[
  colorlinks=true,
  citecolor=blue,
  linkcolor=blue,
  urlcolor=blue
]{hyperref}

\newcommand{\subt}[1]{_\text{#1}}

\newcommand{\norm}[1]{\mleft\lVert #1 \mright\rVert}
\newcommand{\abs}[1]{\mleft\lvert #1 \mright\rvert}
\newcommand{\ii}{\operatorname{i}}
\newcommand{\ee}{\operatorname{e}}
\newcommand{\dd}{\mathrm{d}}

\newcommand{\define}{\mathrel{\mathop:}=}

\newcommand{\Tr}{\operatorname{Tr}}

\newcommand{\poc}[2]{{#1}^{\overline{#2}}}

\NewDocumentCommand{\pFq}{m m m m m}
{
  {}_{#1}F_{#2}\!\left[
    \begin{matrix}
      #3\\
      #4
    \end{matrix}
    ;\,#5
  \right]
}

\begin{document}


\title{Algebraic Characterization of Biphoton Spatial-Mode Entanglement\\ in Higher-Order Laguerre–Gaussian-Pumped SPDC}


\author{Takumi Jinushi}
  \email{takumi.jinushi@gmail.com}
\author{Hirokazu Kobayashi}
 \email{kobayashi.hirokazu@kochi-tech.ac.jp}
\affiliation{
 Graduate School of Engineering,
 Kochi University of Technology,
 185 Miyanokuchi, Tosayamada, Kami City, Kochi 782-8502, Japan
}%


\date{\today}

\begin{abstract}
  High-dimensional spatial entanglement generated via spontaneous parametric down-conversion (SPDC) provides a powerful resource for quantum information processing, yet its mode structure becomes increasingly complex when the pump occupies a higher-order Laguerre--Gaussian (LG) mode.
  Here, we develop an algebraic framework for characterizing biphoton spatial-mode entanglement by factorizing LG-pumped SPDC into a spatial-mode beam-splitter transformation followed by spatial two-mode squeezing.
  By choosing the biphoton LG basis radius as the geometric mean of the pump-beam radius and the crystal-induced correlation width, we obtain a natural modal basis in which the two operations can be treated separately.
  When the pump-beam radius matches the correlation width, spatial squeezing vanishes,
  and the two circular-mode components of the pump are independently conserved across the signal and idler photons, leading to conservation of both orbital angular momentum and the total spatial mode number.
  We further derive an analytical expression for the Schmidt number for arbitrary LG pump modes and squeezing strengths, revealing how the radial and azimuthal pump indices govern the dimensionality of the spatial entanglement.
  This algebraic characterization provides a unified description of the mode structure, conservation laws, and entanglement dimensionality of higher-order-LG-pumped SPDC.

\end{abstract}


\maketitle

\section{\label{sec:intro}Introduction}

Quantum entanglement is a fundamental resource of quantum mechanics and lies at the heart of modern quantum information science~\cite{RevModPhys.81.865,Zhang:24,DENG201746}. Its experimental realization has been firmly established through decades of theoretical and experimental investigations, and today entangled states constitute an essential resource for quantum communication, quantum computation, and quantum sensing. In particular, increasing the dimensionality of quantum states offers a promising route toward enhancing the information capacity and functionality of quantum technologies~\cite{Erhard2020,Forbes2025}. Photonic systems are particularly well suited for this purpose, as multiple degrees of freedom (DoFs) can be exploited to encode high-dimensional quantum states. Among them, the transverse spatial DoFs of light provide a theoretically unbounded Hilbert space and thus offer a powerful platform for realizing high-dimensional quantum states~\cite{suraj2026}.

Spontaneous parametric down-conversion (SPDC) is one of the most widely used nonlinear optical processes for generating entangled photon pairs, and its spatial DoFs provide a rich platform for high-dimensional entanglement. In particular, the Laguerre--Gaussian (LG) mode basis is characterized by two mode indices: the azimuthal mode $l$, associated with orbital angular momentum (OAM) $\hbar l$, and the radial mode $p$, which determines the radial structure of the optical field~\cite{PhysRevA.45.8185,PhysRevA.92.063841,G.Nienhuis2017}. While OAM entanglement associated with $l$ has been extensively investigated, the radial DoF represented by $p$ provides an additional resource for increasing the dimensionality of spatially entangled states~\cite{PhysRevA.83.033816, PhysRevLett.104.020505,PhysRevA.98.042134,Liu:19,Valencia_2021,Forbes2025}.

The spatial correlations generated in SPDC are commonly characterized through the spatial overlap of the pump, signal, and idler modes~\cite{10.10631.1656831,WALBORN201087,PhysRevA.81.053805,PhysRevA.106.063714,Alessio2026}. This approach provides the coupling amplitudes between individual spatial modes~\cite{PhysRevA.102.052412,Yao_2011}, but does not by itself give a simple algebraic picture of how the spatial structure of the pump is distributed among the generated photons. For a fundamental-mode pump, the resulting spatial correlations are well understood in terms of two-mode squeezing, with the generated state exhibiting the structure of a two-mode squeezed vacuum~\cite{PhysRevLett.92.127903,Fedorov_2009}. However, when the pump itself occupies a higher-order spatial mode, it remains unclear how its spatial structure is algebraically distributed between the signal and idler modes and how the resulting multimode entanglement can be described in terms of elementary quantum-optical operations.

Here, we develop a ladder-operator-based algebraic framework for characterizing biphoton spatial-mode entanglement by factorizing higher-order LG-pumped SPDC into an operational sequence of a spatial-mode beam splitter (SBS) followed by a spatial two-mode squeezing operator.
The spatial structure of a higher-order pump beam is coherently redistributed between the signal and idler fields through the beam-splitter-like mode mixing, while the geometric mismatch between the pump beam radius and the crystal-induced correlation width governs the subsequent spatial two-mode squeezing.
This formulation replaces complicated overlap-integral evaluations with an operator-based description of the multimode biphoton state and provides a clear physical picture of how the spatial structure of a higher-order pump mode is transferred to the generated biphotons.

Within this framework, we choose the biphoton LG basis radius as the geometric mean of the pump-beam radius and the crystal-induced correlation width, for which the LG basis coincides with the Schmidt basis in the fundamental-pump case.
In particular, when the pump-beam radius matches the correlation width, the spatial two-mode squeezing vanishes, and the two circular-mode numbers of the pump are independently conserved in the signal-idler pair, leading to the simultaneous conservation of the OAM and the total spatial mode number $N=2p+|l|$ between the pump and the generated photon pair.
Unlike OAM conservation, which follows from rotational symmetry, the conservation of the total spatial mode number emerges only under this matching condition and therefore constitutes an engineered spatial-mode conservation law.
We further derive an analytical expression for the Schmidt number for arbitrary higher-order LG pump modes and squeezing strengths.
Together, these results provide a unified algebraic characterization of the spatial-mode structure, conservation properties, and entanglement dimensionality of higher-order-LG-pumped SPDC.

The remainder of this paper is organized as follows. In Sec.~\ref{sec:main}, we present the mathematical formulation of LG beams and describe the SPDC process in the transverse wavevector space.
In Sec.~\ref{sec:sbs}, we introduce the operational decomposition of the LG-pumped SPDC process, factoring the interaction into a spatial-mode beam splitter and a spatial two-mode squeezing operator. Sec.~\ref{sec:LGmode_decom} discusses the spatial-mode conservation laws, highlighting the strict conservation of the total mode number $N$, and examines the spatial-mode decomposition under different squeezing regimes. In Sec.~\ref{sec:sch-num}, we quantitatively analyze the Schmidt number $K$ to evaluate the dimensionality of the generated spatial entanglement. Finally, we draw our conclusions in Sec.~\ref{sec:conc}. The detailed mathematical derivations of the operational transformations, the spatial-mode distributions using difference operators, and the Schmidt numbers via generating functions are provided in Appendix~\ref{app:squeezing}, \ref{app:algebraic}, and \ref{app:schmidt}, respectively.

\section{\label{sec:main}Laguerre--Gaussian beams and spontaneous parametric down-conversion}

An LG mode state $\ket{N_+,N_-}$, where $N_\pm$ are the circular-mode numbers, is generated by applying the circular raising operators $\hat{a}^\dagger_\pm$, which shift the OAM by $\pm\hbar$ and increase the total mode number by one~\cite{G.Nienhuis2017}, to the fundamental mode $\ket{0,0}$:
\begin{align} \label{eq:LGmode;main}
  \ket{N_+,N_-}
  &\define {\hat{a}_{N_+,N_-}}^\dagger \! \ket{0,0} \\
  &\define C_{N_+}^{N_-} ({\hat{a}_+}^\dagger)^{N_+} ({\hat{a}_-}^\dagger)^{N_-} \ket{0,0},
\end{align}
where ${\hat{a}_{N_+,N_-}}^\dagger$ denotes the spatial-mode raising operator~\cite{PhysRevA.67.052313,PhysRevA.98.043824,Hiekkamaki2022}, ${(\cdot)}^\dagger$ is the Hermitian conjugate, and $C_{N_+}^{N_-} \! \define \! 1/\sqrt{N_+!N_-!}$ is the normalization coefficient~\cite{PhysRevA.48.656}.
The total mode number is $N \define N_+ + N_- = 2p+\abs{l}$ where $p \!\define\! \min(N_+, N_-)$ is the radial mode index and $l \define N_+ - N_-$ is the azimuthal mode index~\cite{PhysRevA.48.656,PhysRevA.89.063813}. To describe the physical observables corresponding to these modes, the mode number operators $\hat{N}_\pm$, the dimensionless OAM operator $\hat{L}_z$, and the total mode number operator $\hat{N}$ are constructed from the ladder operators $\{{\hat{a}_{\pm}},{\hat{a}_{\pm}}^\dagger\}$ as follows~\cite{PhysRevA.48.656,PhysRevA.89.063813}:
\begin{gather}
  \hat{N}_\pm
  \define \hat{a}_\pm^\dagger \hat{a}_\pm, \
  \hat{L}_z
  \define \hat{N}_+ - \hat{N}_-, \
  \hat{N}
  \define \hat{N}_+ + \hat{N}_-.
\end{gather}
In the wavevector representation, an LG beam $\ket{N_+,N_-}$ is expressed as~\cite{PhysRevA.45.8185,PhysRevLett.119.263602}
\begin{align} \label{eq:LGbeam;main}
  \text{LG}_{N_+}^{N_-} \! (\bold{k}) \!
  &\define \braket{\bold{k}|N_+,N_-} \nonumber \\
  &=\! C_p^l \mleft[\dfrac{w\subt{p} k_r}{\sqrt{2}} \mright]^{\abs{l}}
    L_p^{\abs{l}} \! \mleft[\dfrac{w\subt{p}^2 {k_r}^2}{2}\mright]
    \! \ee^{-{w\subt{p}}^2 {k_r}^2 \! / 4}
    \ee^{\ii \! l k_\theta}\!,
\end{align}
where $(k_r,k_\theta)$ are the polar coordinates of the transverse wavevector $\bold{k}$, the beam radius is $w\subt{p}$, $L_p^{\abs{l}}$ is the associated Laguerre polynomial, and the normalization coefficient $C_p^l$ is
\begin{align}
  C_p^l
  &\define (-\! \ii)^{\abs{l}} \pi w\subt{p} \sqrt{ \dfrac{2}{\pi}\! \dfrac{p!}{(p+\abs{l})!}}.
\end{align}

In the SPDC process, a pump field interacting with a second-order nonlinear optical crystal (NLC) generates correlated signal and idler photons.
Assuming a weak interaction between the classical pump field and the NLC, with a propagation time $T$ through the NLC, the biphoton state $\ket{\Psi_{N_+}^{N_-}}$ is generated from the vacuum state $\ket{\text{vac}}$ via the second-order nonlinear interaction Hamiltonian $\hat{H}\subt{SPDC}$ as follows:
\begin{align} \label{eq:biphoton-Hamiltonian;main}
  \ket{\Psi_{N_+}^{N_-}}
  \approx -\dfrac{\ii}{\hbar} \int_{0}^{T} \odif{t} ~
    \hat{H}\subt{SPDC} \ket{\text{vac}}.
\end{align}
For simplicity, we assume the pump light to be a monochromatic continuous wave (CW) with a wavelength of $\lambda\subt{p}$ and to be collimated. The generated biphotons are assumed to be collinear at the degenerate wavelength $\lambda\subt{s,i} \define 2\lambda\subt{p}$, where the subscripts p, s, and i denote the pump, signal, and idler photons, respectively.

Under these conditions, the second-order nonlinear interaction Hamiltonian is expressed in the transverse wavevector space as~\cite{PhysRevA.95.063836,WALBORN201087,Fedorov_2009,PhysRevA.106.063714}:
\begin{align} \label{eq:hamiltonian;main}
  \hat{H}\subt{SPDC}
  &\!\propto\!
    \int \!\!\dfrac{\!\odif{\bold{k}\subt{s}}\!}{(2\pi)^2} \!
    \int \!\!\dfrac{\!\odif{\bold{k}\subt{i}}\!}{(2\pi)^2}
    \text{LG}_{N_+}^{N_-}(\bold{k}\subt{s}+\bold{k}\subt{i}) \!
    \nonumber \\ & \quad \times
    \exp\mleft[
      - \dfrac{{w_-}^2}{4} \norm{\bold{k}\subt{s}-\bold{k}\subt{i}}^2
    \mright]
    {\hat{a}\subt{s}}^\dagger(\bold{k}\subt{s})
    {\hat{a}\subt{i}}^\dagger(\bold{k}\subt{i})
  + \text{H.c.},
\end{align}
where $\text{LG}_{N_+}^{N_-}(\bold{k}\subt{p})$ is the complex amplitude of the LG pump field and $w_- \define \sqrt{0.718 \lambda\subt{p} L/2\pi}$, with $L$ being the NLC length, denotes the crystal-induced correlation width under the double-Gaussian approximation~\cite{PhysRevA.106.063714}, H.c. denotes the Hermitian conjugate, and we have used the transverse-momentum conservation $\hbar\bold{k}\subt{p} = \hbar\bold{k}\subt{s} + \hbar\bold{k}\subt{i}$ and the energy conservation $\hbar\omega\subt{p} = \hbar\omega\subt{s} + \hbar\omega\subt{i}$.
The generated biphoton state $\ket{\Psi_{N_+}^{N_-}}$ is obtained by applying the Hamiltonian in Eq.~\eqref{eq:hamiltonian;main} to the vacuum state according to Eq.~\eqref{eq:biphoton-Hamiltonian;main}.
To connect this continuous-wavevector description with the algebraic spatial-mode representation developed below, we expand the biphoton state $\ket{\Psi_{N_+}^{N_-}}$ in the discrete LG-mode basis $\ket{N_{\text{s}+}, N_{\text{s}-}}\subt{s} \ket{N_{\text{i}+}, N_{\text{i}-}}\subt{i}$:
  \begin{align}
    \ket{\Psi_{N_+}^{N_-}}
    &\!=\! \sum_{N_{\text{s}+},N_{\text{s}-}}
      \sum_{N_{\text{i}+},N_{\text{i}-}}
      c_{N_{\text{s}+},N_{\text{s}-}}^{N_{\text{i}+},N_{\text{i}-}}
      \ket{N_{\text{s}+}, N_{\text{s}-}}\subt{s}
      \ket{N_{\text{i}+}, N_{\text{i}-}}\subt{i},
\end{align}
where the spatial-mode decomposition coefficients are defined as $c_{N_{\text{s}+},N_{\text{s}-}}^{N_{\text{i}+},N_{\text{i}-}} \define \bra{N_{\text{s}+}, N_{\text{s}-}}\!\braket{N_{\text{i}+}, N_{\text{i}-}|\Psi_{N_+}^{N_-}}$.
In general, these coefficients depend on the beam radius chosen for the LG basis, so that different basis radii yield different modal decompositions of the same physical biphoton state.
In the following section, we show that the operational decomposition of the SPDC spatial transformation provides a natural choice for the basis radius.

\section{\label{sec:sbs}Operational decomposition of LG-pumped SPDC}

\begin{figure}
  \centering
  \includegraphics[width=.5\textwidth]{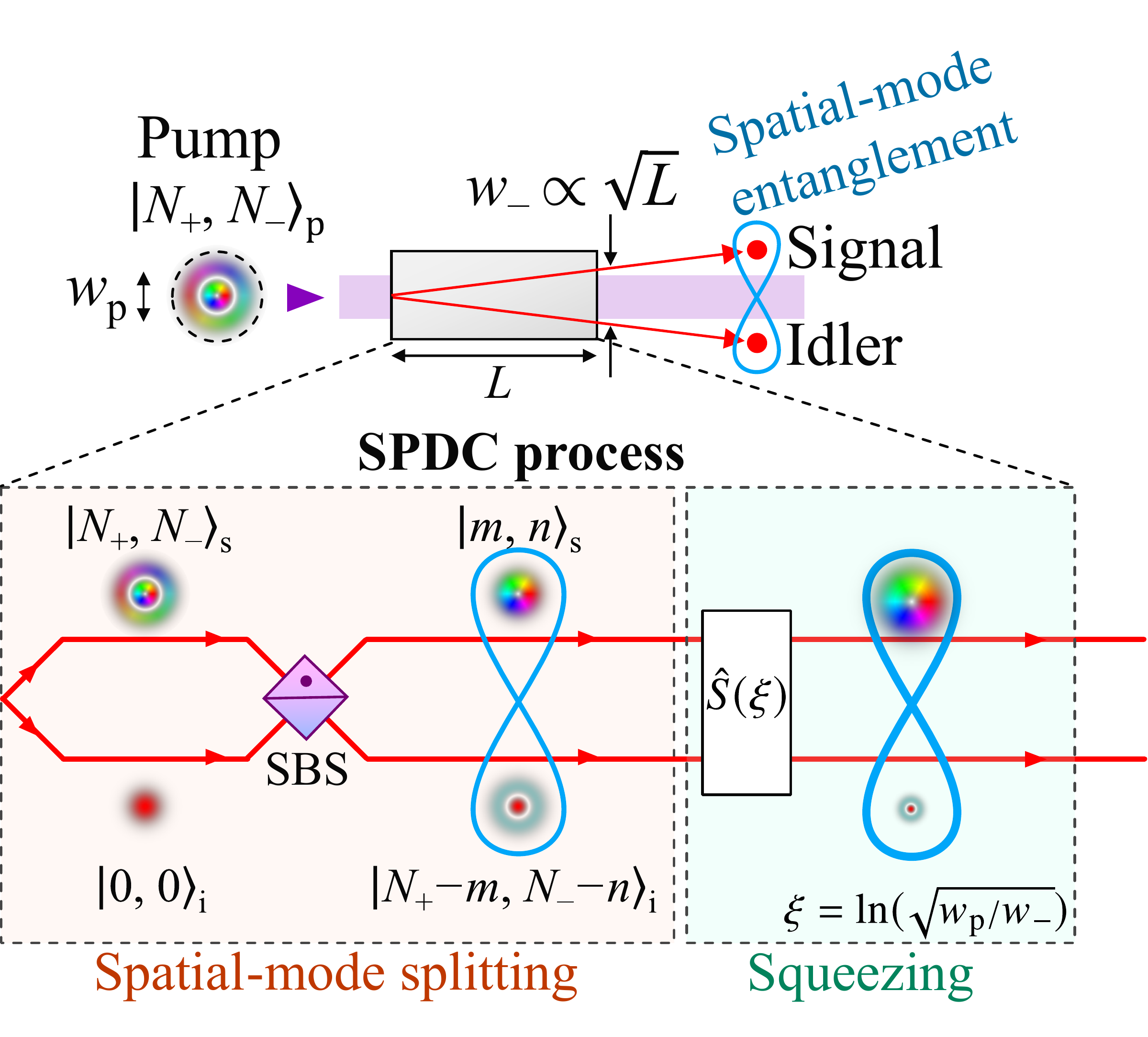}
  \caption{
    Schematic of the operational decomposition of LG-pumped SPDC.
    The spatial-mode beam-splitter transformation $\hat U_{\rm SBS}$ coherently redistributes the pump-mode excitations between the signal and idler, after which the spatial two-mode squeezing operator $\hat S(\xi)$ acts on the resulting biphoton state.
    The squeezing parameter $\xi=\ln\sqrt{w\subt{p}/w_-}$ is determined by the mismatch between the pump-beam radius $w\subt{p}$ and the crystal-induced correlation width $w_-$.
    Under the correlation-width matching condition $w\subt{p}=w_-$, spatial squeezing vanishes $(\xi=0)$.
  }
  \label{fig:SPDC_SBS-rep;sbs}
\end{figure}

From Eqs.~\eqref{eq:biphoton-Hamiltonian;main} and \eqref{eq:hamiltonian;main}, the SPDC biphoton amplitude in transverse-wavevector space, $\braket{\bold{k}\subt{s},\bold{k}\subt{i}|\Psi_{N_+}^{N_-}}$, is naturally expressed in terms of the sum and difference coordinates, $\bold{k}_+\define(\bold{k}_s+\bold{k}_i)/\sqrt{2}$ and $\bold{k}_-\define(\bold{k}_s-\bold{k}_i)/\sqrt{2}$, associated with the pump spatial structure and the crystal-induced spatial correlation, respectively.
This structure can be interpreted as a combination of two distinct transformations: beam-splitter-like mixing of the signal and idler spatial modes and relative spatial scaling along the two collective coordinates $(\bold{k}_+, \bold{k}_-)$.
The former corresponds to the linear transformation from $(\bold{k}\subt{s},\bold{k}\subt{i})$ to $(\bold{k}_+,\bold{k}_-)$, while the latter can be represented as spatial two-mode squeezing by choosing the LG-basis reference radius as
\begin{align} \label{eq:w0;sbs}
w_0:=\sqrt{2w\subt{p} w_-},
\end{align}
where $w\subt{p}$ is the pump beam radius and $w_-$ is the correlation width.
For a fundamental Gaussian pump, Eq.~\eqref{eq:w0;sbs} makes the LG basis coincide with the genuine Schmidt basis of the biphoton state.
With this choice, the biphoton spatial transformation can be factorized into a spatial-mode beam-splitter transformation followed by spatial two-mode squeezing, as shown in Fig.~\ref{fig:SPDC_SBS-rep;sbs}. The resulting biphoton state is given by
\begin{align} \label{eq:biphoton-entangle-sd-LGlp-N+-;sbs}
  \ket{\Psi_{N_+}^{N_-}(\xi)} \!
  &= \hat{S}(\xi) \hat{U}\subt{SBS}
    \ket{N_+,N_-}\subt{s} \ket{0,0}\subt{i},
\end{align}
where $N_+$ and $N_-$ are the circular-mode numbers of the pump field, $\hat{U}\subt{SBS}$ is the spatial-mode beam-splitter operator, and $\hat{S}(\xi)$ is the spatial two-mode squeezing operator with the squeezing parameter $\xi$ determined by the transverse-scale mismatch between the pump beam radius $w\subt{p}$ and the correlation width $w_-$:
\begin{align} \label{eq:xi;sbs}
  \xi \define \ln(\sqrt{w\subt{p}/w_-}).
\end{align}
The complete derivations supporting this algebraic framework are detailed in Appendix~\ref{app:squeezing}.

The SBS operator $\hat{U}_{\text{SBS}}$ provides an algebraic description of the coherent redistribution of the pump-mode excitations between the signal and idler spatial modes.
It acts on the signal and idler spatial-mode raising operators according to~\cite{PhysRevLett.45.75}:
\begin{align} \label{eq:sbs;sbs}
  \begin{aligned}
  \hat{U}_{\text{SBS}}
  \hat{a}_{\text{s}\pm}^\dagger
  \hat{U}_{\text{SBS}}^\dagger
  &\!=\! \dfrac{\hat{a}_{\text{s}\pm}^\dagger\!\!+\!\hat{a}_{\text{i}\pm}^\dagger}{\sqrt{2}},
  \\
  \hat{U}_{\text{SBS}}
  \hat{a}_{\text{i}\pm}^\dagger
  \hat{U}_{\text{SBS}}^\dagger
  &\!=\! \dfrac{\hat{a}_{\text{s}\pm}^\dagger\!\!-\!\hat{a}_{\text{i}\pm}^\dagger}{\sqrt{2}}. \!
  \end{aligned}
\end{align}
Although its algebraic structure resembles that of a conventional 50:50 optical beam splitter, $\hat{U}\subt{SBS}$ does not represent a physical beam splitter acting on distinct optical paths, but instead acts on the signal-idler spatial-mode degrees of freedom.

The remaining relative spatial scaling is described by the spatial two-mode squeezing operator~\cite{gu2026analyticalfockrepresentationtwomode}
\begin{align} \label{eq:squeez;sbs}
  \hat{S}(\xi)
  \define \ee^{\xi(\hat{X}^\dagger-\hat{X})},
\end{align}
where the bipartite operator is defined as $\hat{X} \define \hat{a}_{\text{s}+}\hat{a}_{\text{i}-} + \hat{a}_{\text{s}-}\hat{a}_{\text{i}+}$.
Under the correlation-width matching condition $w\subt{p}=w_-$, the squeezing parameter vanishes, $\xi=0$, and $\hat{S}(0)$ reduces to the identity operator; the biphoton spatial structure is then determined solely by the SBS mode mixing.
The cross-coupled form of $\hat{X}$ shows that the squeezing acts between opposite circular spatial-mode components of the signal and idler, with the two terms representing mutually commuting two-mode squeezing processes.
The commutation relation
\begin{align}
  [\hat{X},\hat{X}^\dagger]=\hat{N}+2,
\end{align}
where
$\hat{N}\define
\hat{N}\subt{s}+\hat{N}\subt{i}
=\hat{N}_{\text{s}+}+\hat{N}_{\text{s}-}
+\hat{N}_{\text{i}+}+\hat{N}_{\text{i}-}$
is the total spatial mode number operator of the signal-idler system, characterizes the algebraic structure of this spatial squeezing transformation.
The interplay between the SBS transformation and spatial squeezing can also be seen directly from the transformation of the pair-annihilation operator,
\begin{align}
  \hat{U}\subt{SBS}^\dagger \hat{X} \hat{U}\subt{SBS}
  &=\hat{a}_{\text{s}+} \hat{a}_{\text{s}-}
  - \hat{a}_{\text{i}+} \hat{a}_{\text{i}-}.
\end{align}
Thus, in the SBS-transformed basis, the cross-coupled signal--idler pair creation is mapped onto opposite-circular-mode pair creation within the signal and idler subsystems.

\section{\label{sec:LGmode_decom}Spatial-mode structure and conservation in LG-pumped SPDC}

\begin{figure}
  \centering
  \includegraphics[width=.5\textwidth]{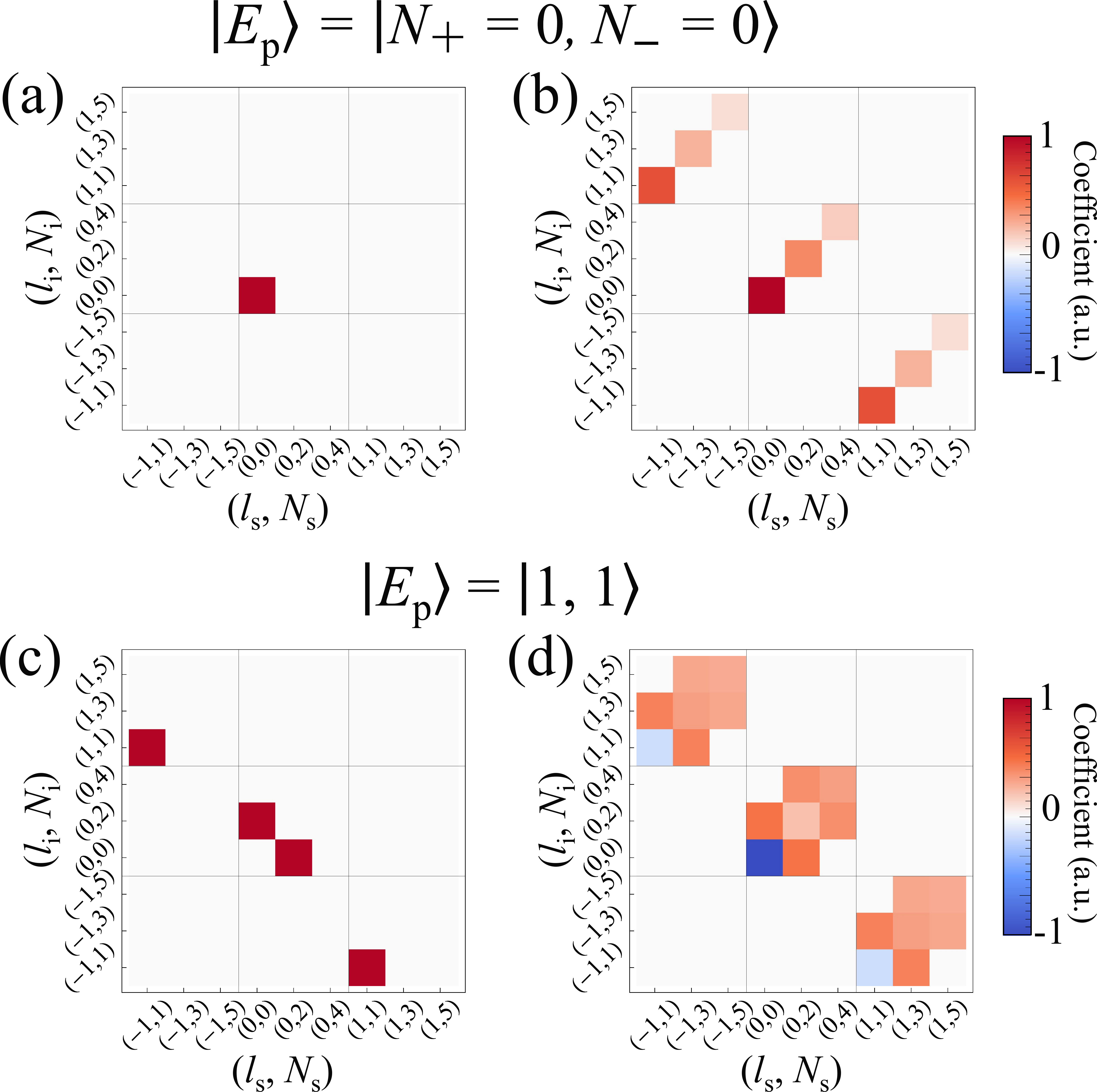}
  \caption{
    Biphoton spatial-mode distributions in the LG basis with reference radius $w_0$ for a fundamental Gaussian pump $\ket{0,0}\subt{p}$ (top row) and a higher-order LG pump $\ket{1,1}\subt{p}$ (bottom row).
    (a), (c) Under the correlation-width matching condition $\xi=0$, spatial squeezing vanishes and the distributions are confined to a finite mode manifold satisfying $N\subt{s}+N\subt{i}=N$.
    (b), (d) For $\xi\neq0$, spatial squeezing broadens the distributions beyond this fixed-total-mode-number manifold while preserving the finite mode-difference range $-N\leq\Delta N\subt{s,i}\leq N$.
    For the fundamental pump in (b), $\Delta N\subt{s,i}=0$, yielding one-to-one diagonal pairing between conjugate signal and idler modes.
  }
  \label{fig:LG01_SD;LGmode_decom}
\end{figure}

We now examine how the spatial-mode distribution of the biphoton state in Eq.~\eqref{eq:biphoton-entangle-sd-LGlp-N+-;sbs} and the associated mode-conservation properties depend on both the pump-mode structure and the spatial squeezing parameter $\xi$.
Figure~\ref{fig:LG01_SD;LGmode_decom} provides an overview of the biphoton spatial-mode distributions in the LG basis with the reference radius $w_0$ defined in Eq.~\eqref{eq:w0;sbs}. The fundamental and higher-order LG pump modes, $(N_+,N_-)=(0,0)$ and $(1,1)$, respectively, are compared under the correlation-width matching condition $\xi=0$ and in a spatially squeezed regime $\xi=+0.74$.
For $\xi=0$, spatial squeezing is absent, and the biphoton mode distributions in Figs.~\ref{fig:LG01_SD;LGmode_decom}(a) and (c) are confined to a finite set of LG modes constrained by the pump-mode structure and the associated conservation laws discussed below.
When spatial squeezing is present $(\xi\neq0)$, on the other hand, the distributions in Figs.~\ref{fig:LG01_SD;LGmode_decom}(b) and (d) broaden beyond the finite mode sets obtained at $\xi=0$, while preserving OAM conservation.
For a fundamental Gaussian pump, the broadened distribution retains a one-to-one pairing between conjugate signal and idler modes, whereas for a higher-order pump, the initial mode distribution generated by the SBS transformation is further broadened by spatial squeezing.
The correlation-width matching ($\xi=0$) and spatial-squeezing regimes  ($\xi\neq 0$) are analyzed separately in the following subsections.

\subsection{Correlation-width matching condition: $\xi=0$}

Under the correlation-width matching condition $w\subt{p}=w_-$, the squeezing parameter vanishes, $\xi=0$, and the spatial two-mode squeezing operator reduces to the identity, $\hat{S}(0)=\hat{I}$.
The biphoton state in Eq.~\eqref{eq:biphoton-entangle-sd-LGlp-N+-;sbs} is therefore determined solely by the SBS transformation,
\begin{align}
\ket{\Psi_{N_+}^{N_-}(0)}
=
\hat{U}\subt{SBS}
\ket{N_+,N_-}\subt{s}\ket{0,0}\subt{i}.
\end{align}
Using the beam-splitter transformation of the spatial-mode raising operators in Eq.~\eqref{eq:sbs;sbs}, this state can be expanded as follows~\cite{PhysRevLett.45.75}:
\begin{align} \label{eq:UBSpsi;LGmode_decom}
  \ket{\Psi_{N_+}^{N_-}(0)}
  &= \sum_{m=0}^{N_+} \sum_{n=0}^{N_-}
    \sqrt{\dfrac{1}{2^N} \binom{N_+}{m} \binom{N_-}{n}} \nonumber \\ & \quad \times
    \ket{m,n}\subt{s}
    \ket{N_+-m,N_--n}\subt{i},
\end{align}
where $\binom{n}{k}=n!/[k!(n-k)!]$ denotes the binomial coefficient.
Thus, the pump-mode excitations are coherently distributed between the signal and idler while preserving the two circular-mode numbers independently:
\begin{align} \label{eq:N+,N-;LGmode_decom}
  \begin{aligned}
N_+ &= N_{\text{s}+}+N_{\text{i}+},\\
N_- &= N_{\text{s}-}+N_{\text{i}-}.
\end{aligned}
\end{align}
A central consequence of our algebraic framework is that these two independent selection rules imply not only the usual conservation of the azimuthal mode number but also conservation of the total spatial mode number:
\begin{align} \label{eq:Lz-N_cons;LGmode_decom}
  \begin{aligned}
l &= l\subt{s}+l\subt{i},\\
N &= N\subt{s}+N\subt{i}.
\end{aligned}
\end{align}
To the best of our knowledge, this total spatial-mode-number conservation has not previously been identified in SPDC.
While OAM conservation follows from the rotational symmetry of the SPDC interaction, conservation of the total spatial mode number appears only under the correlation-width matching condition and restricts the OAM spectral bandwidth to a finite range~\cite{PhysRevLett.104.020505}.
It therefore represents an emergent mode-conservation property associated with the suppression of spatial squeezing.

The finite mode distributions shown in Figs.~\ref{fig:LG01_SD;LGmode_decom}(a) and (c) follow directly from these constraints.
For the fundamental Gaussian pump, $(N_+,N_-)=(0,0)$, no spatial-mode correlation remains under the correlation-width matching condition~\cite{Paul_2018}, and the biphoton state reduces to the separable product state
\begin{align} \label{eq:no-corr;LGmode_decom}
\ket{\Psi_0^0(0)}
=\ket{0,0}\subt{s}\ket{0,0}\subt{i},
\end{align}
as shown in Fig.~\ref{fig:LG01_SD;LGmode_decom}(a).
For a higher-order LG pump, by contrast, spatial entanglement persists even at $\xi=0$, because the pump-mode excitations are coherently distributed between the signal and idler by the SBS transformation.
Consequently, the biphoton state forms a finite superposition of LG modes with the total mode number fixed by the pump order $N=N_++N_-$~\cite{PhysRevLett.104.020505}, as illustrated in Fig.~\ref{fig:LG01_SD;LGmode_decom}(c).
In this regime, the LG-mode expansion in Eq.~\eqref{eq:UBSpsi;LGmode_decom} constitutes the Schmidt decomposition of the biphoton state, so that the remaining spatial entanglement originates entirely from the higher-order spatial structure of the pump rather than from spatial squeezing.

\subsection{Spatial squeezing regime: $\xi\neq 0$}

When the pump beam radius and the crystal-induced correlation width are not matched, $\xi\neq0$, the spatial two-mode squeezing operator acts nontrivially on the finite SBS-generated state discussed in the previous subsection.
Consequently, the individual circular-mode-number conservation laws in Eq.~\eqref{eq:N+,N-;LGmode_decom} no longer hold, and the biphoton distribution is not restricted to a fixed total spatial mode number $N\subt{s}+N\subt{i}=N$.

Nevertheless, in the LG basis with the reference radius $w_0$, the squeezing transformation preserves specific differences between the signal and idler circular-mode occupations, defined as
\begin{align}
  \begin{aligned}
\Delta\hat{N}_{\text{s}+,\text{i}-}
&\define\hat{N}_{\text{s}+}-\hat{N}_{\text{i}-},\\
\Delta\hat{N}_{\text{s}-,\text{i}+}
&\define\hat{N}_{\text{s}-}-\hat{N}_{\text{i}+}.
\end{aligned}
\end{align}
These operators commute with the spatial two-mode squeezing operator,
\begin{align} \label{eq:[S,N_Delta];LGmode_decom}
  [\hat{S}(\xi),\Delta\hat{N}_{\text{s}+,\text{i}-}]
  = [\hat{S}(\xi),\Delta\hat{N}_{\text{s}-,\text{i}+}]
  =0.
\end{align}
Thus, although spatial squeezing generates excitations over an unbounded range of total mode number, it does so within mode-difference sectors fixed by the SBS-generated seed state.
The total OAM operator of the signal-idler system can be written as
\begin{align}
\hat{L}\subt{sz}+\hat{L}\subt{iz}
=
\Delta\hat{N}_{\text{s}+,\text{i}-}
-\Delta\hat{N}_{\text{s}-,\text{i}+}.
\end{align}
It therefore also commutes with the spatial two-mode squeezing operator $\hat{S}(\xi)$, and OAM conservation is retained even in the presence of spatial squeezing.

For a fundamental Gaussian pump, $(N_+,N_-)=(0,0)$, the biphoton state with $\xi\neq0$ is generated by applying spatial two-mode squeezing to the state $\ket{\psi_0^0(0)}$ in Eq.~\eqref{eq:no-corr;LGmode_decom}:
\begin{align} \label{eq:biphoton-sd-LG00;LGmode_decom}
\ket{\Psi_0^0(\xi)}
=
\hat{S}(\xi)
\ket{0,0}\subt{s}
\ket{0,0}\subt{i}.
\end{align}
In this case, the initial values of both mode differences are zero,
$\Delta N_{\text{s}+,\text{i}-}=\Delta N_{\text{s}-,\text{i}+}=0$, and therefore
\begin{align}
N_{\text{s}\pm}=N_{\text{i}\mp}.
\end{align}
The spatial-mode distribution consequently broadens while maintaining a one-to-one pairing between conjugate signal and idler LG modes, as shown in Fig.~\ref{fig:LG01_SD;LGmode_decom}(b).
Accordingly, in the LG basis with the reference radius $w_0$, the mode expansion remains the Schmidt decomposition of the biphoton state for a fundamental Gaussian pump even when $\xi\neq0$~\cite{PhysRevLett.92.127903,Fedorov_2009,Miatto2012}.

For a higher-order LG pump, by contrast, the SBS transformation first produces a finite superposition of states with different signal--idler mode differences.
The subsequent squeezing operation broadens each of these components while preserving its mode-difference quantum numbers.
As a result, the strict one-to-one pairing characteristic of the fundamental Gaussian pump is lost, and the LG basis generally no longer constitutes the Schmidt basis, as illustrated in Fig.~\ref{fig:LG01_SD;LGmode_decom}(d).
Nevertheless, because
$\Delta\hat{N}\subt{s,i}\define
\Delta\hat{N}_{\text{s}+,\text{i}-}
+\Delta\hat{N}_{\text{s}-,\text{i}+}$
is conserved by the squeezing operation, the squeezed biphoton state remains confined within
\begin{align}
-N\leq \Delta N\subt{s,i} \leq N.
\end{align}
Thus, spatial squeezing removes the fixed-total-mode-number constraint found at $\xi=0$ while leaving well-defined mode-difference constraints that determine the structure of the higher-order biphoton spatial-mode distribution.
The explicit analytical expression for the resulting squeezed higher-order biphoton state is given in Appendix~\ref{app:algebraic}.

\section{\label{sec:sch-num}Dimensionality of biphoton spatial entanglement}

As shown in Sec.~\ref{sec:LGmode_decom}, the spatial-mode distribution of the SPDC biphoton is confined to a finite mode manifold under the correlation-width matching condition $\xi=0$, whereas spatial squeezing for $\xi\neq0$ broadens the distribution over an increasing number of LG modes.
We now quantify the dimensionality of the resulting biphoton entanglement through the Schmidt number, $K$.

As a reference, we first consider the fundamental Gaussian pump, for which the Schmidt number has the well-known analytical dependence on the squeezing parameter $\xi$~\cite{Fedorov_2009,PhysRevLett.92.127903,PhysRevA.69.052117}:
\begin{align} \label{eq:K00;sch-num}
  K_0^0(\xi)=\cosh^2(2\xi).
\end{align}
This expression is an even function of $\xi$ and reaches its global minimum, $K_0^0(0)=1$, at the correlation-width matching condition $\xi=0$.
Under this condition, $w\subt{p}=w_-$ and $w\subt{s,i}=w_0=\sqrt{2}w\subt{p}$, spatial squeezing vanishes, and the biphoton spatial state reduces to the separable product of fundamental Gaussian modes in Eq.~\eqref{eq:no-corr;LGmode_decom}~\cite{Paul_2018}.
For sufficiently large $\abs{\xi}$, by contrast, the Schmidt number approaches the asymptotic exponential scaling
\begin{align} \label{eq:K00-asym;sch-num}
  K_0^0(\xi) \approx \frac{1}{4}\ee^{4\abs{\xi}},
\end{align}
showing that the entanglement dimensionality increases exponentially with increasing squeezing magnitude $\abs{\xi}$.

We next extend the analysis to higher-order LG pump modes, beginning with the correlation-width matching condition $\xi=0$.
Unlike the fundamental Gaussian pump case where $K_0^0(0)=1$, a higher-order pump generates spatial entanglement even in the absence of spatial squeezing.
As shown in Sec.~\ref{sec:LGmode_decom}, the SBS transformation coherently distributes the pump-mode excitations between the signal and idler while satisfying the mode-conservation constraints, thereby confining the biphoton state to a finite spatial-mode manifold.
Under this condition, Eq.~\eqref{eq:UBSpsi;LGmode_decom} is already in Schmidt form, with Schmidt eigenvalues
\begin{align}
\lambda_{m,n}
=
\frac{1}{2^N}
\binom{N_+}{m}
\binom{N_-}{n}.
\end{align}
The Schmidt number can therefore be evaluated directly as
\begin{align} \label{eq:K-xi=0;sch-num}
K_{N_+}^{N_-}(0)
&=
\left(
\sum_{m=0}^{N_+}
\sum_{n=0}^{N_-}
{\lambda_{m,n}}^2
\right)^{-1}
\nonumber\\
&=
4^N
\binom{2N_+}{N_+}^{-1}
\binom{2N_-}{N_-}^{-1}.
\end{align}
For highly excited pump modes with both $N_+$ and $N_-$ large, the combinatorial expression in Eq.~\eqref{eq:K-xi=0;sch-num} can be evaluated asymptotically.
Using Stirling's formula~\cite{elezovic2014asymptotic}, we obtain
\begin{align} \label{eq:K-xi=0_approx;sch-num}
  K_{N_+}^{N_-}(0)
  &\approx \pi
    \sqrt{
      \mleft(N_+ + \frac{1}{4} \mright)
      \mleft(N_- + \frac{1}{4} \mright)} \\
  &= \pi
    \sqrt{
      \mleft(p + \frac{1}{4} \mright)
      \mleft(p+\abs{l} + \frac{1}{4} \mright)}.
      \nonumber
\end{align}
This asymptotic expression reveals distinct contributions of the radial and azimuthal pump indices to the mode-conserving entanglement dimensionality.
For fixed $\abs{l}$ and large $p$, the Schmidt number increases approximately linearly, $K\propto p$.
In contrast, for fixed $p$ and large $\abs{l}$, the exact expression in Eq.~\eqref{eq:K-xi=0;sch-num} instead yields $K\propto\sqrt{\abs{l}}$.
Thus, the Schmidt number exhibits a stronger asymptotic dependence on radial excitation than on azimuthal excitation.

We now evaluate the Schmidt number for an arbitrary LG pump mode and squeezing parameter.
The resulting expression can be written as a finite weighted sum over the radial index:
\begin{align}  \label{eq:K-xi;sch-num}
  K_{N_+}^{N_-} (\xi)
  &= K_0^0(\xi) \mleft(
    \sum_{n=0}^{p}
    \mleft[
      K_{N_+-n}^{N_--n}(0)
      K_{n}^{n}(0)
    \mright]^{-1}
    {\mu_2}^{4n}
  \mright)^{-1},
\end{align}
where $\mu_2\define\tanh(2\xi)$.
This closed-form expression fully characterizes the spatial entanglement for arbitrary squeezing strengths and separates two physically distinct contributions to the entanglement dimensionality.
The prefactor $K_0^0(\xi)$ represents the contribution of spatial squeezing already present for a fundamental Gaussian pump in Eq.~\eqref{eq:K00;sch-num}, whereas the finite weighted sum describes the interplay between the higher-order pump structure and the squeezing transformation.
Within each term of the sum, the two factors are mode-conserving Schmidt numbers given by Eq.~\eqref{eq:K-xi=0;sch-num}.
Specifically, $K_{N_+-n}^{N_--n}(0)$ corresponds to a component that retains the pump OAM, since $(N_+-n)-(N_--n)=l$, whereas $K_n^n(0)$ corresponds to a zero-OAM component.
At $\xi=0$, only the $n=0$ term survives, and Eq.~\eqref{eq:K-xi;sch-num} consistently reduces to the mode-conserving result $K_{N_+}^{N_-}(0)$ in Eq.~\eqref{eq:K-xi=0;sch-num}.
Details of the derivation of Eq.~\eqref{eq:K-xi;sch-num} and its alternative formulations are given in Appendix~\ref{app:schmidt}.

\begin{figure}[t]
  \centering
  \includegraphics[width=.5\textwidth]{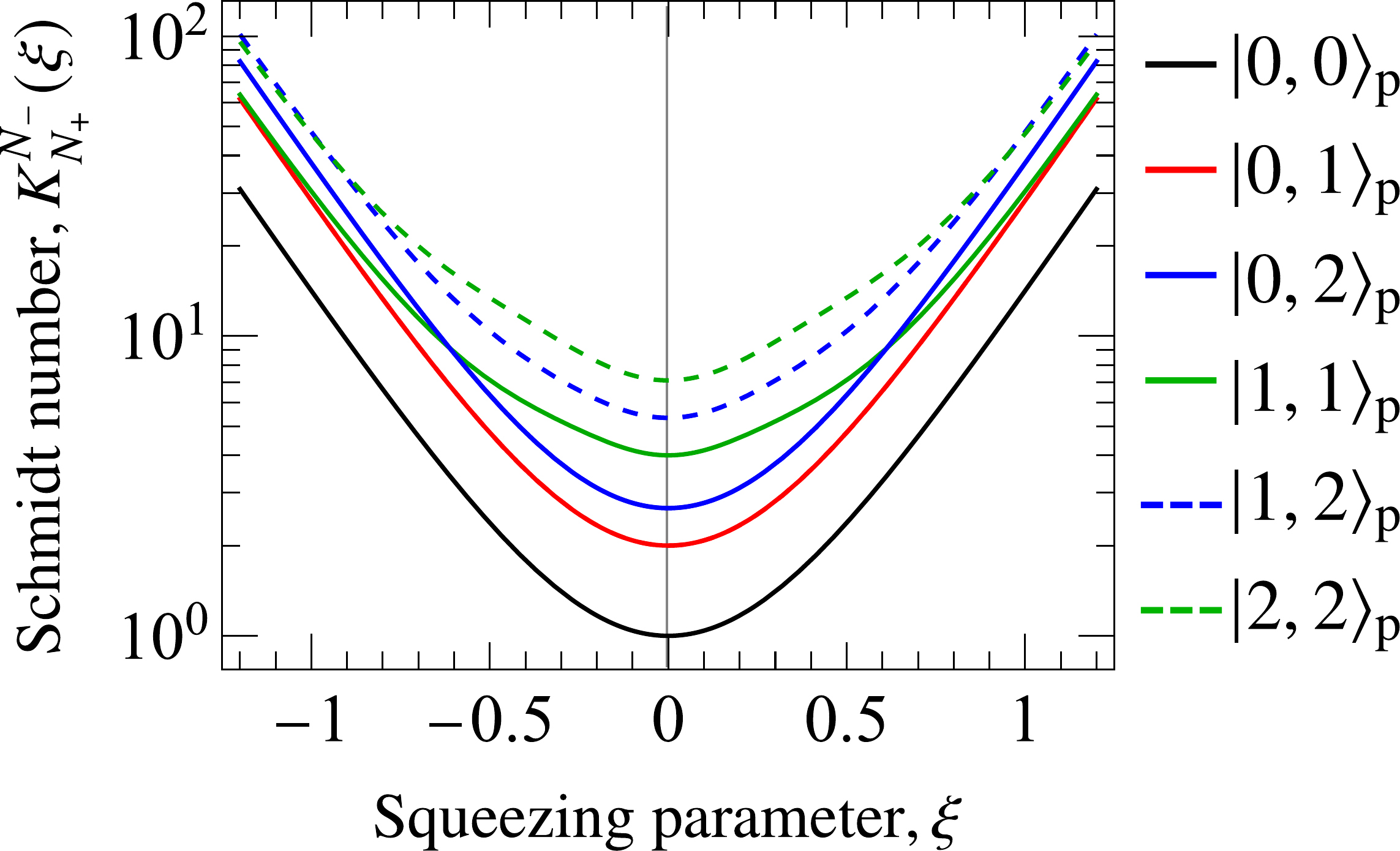}
  \caption{
    Schmidt number $K_{N_+}^{N_-}(\xi)$ as a function of the squeezing parameter $\xi$ for representative LG pump modes $\ket{N_+,N_-}\subt{p}$.
    For all pump modes, the Schmidt number is an even function of $\xi$, reaches its minimum at the correlation-width matching condition $\xi=0$, and increases with $\abs{\xi}$.
  }
  \label{fig:gamma-SN-general;sch-num}
\end{figure}

Equation~\eqref{eq:K-xi;sch-num} shows that the Schmidt number is an even function of $\xi$ and increases monotonically with $\abs{\xi}$, attaining its global minimum at the correlation-width matching condition $\xi=0$, because Eq.~\eqref{eq:K-xi=0;sch-num} is always positive and ${\mu_2}^4$ is an even function of $\xi$. 
Figure~\ref{fig:gamma-SN-general;sch-num} illustrates this behavior for representative LG pump modes.
The detailed dependence on $\xi$, however, varies with the spatial structure of the pump.
We therefore examine the weak- and strong-squeezing limits separately below.

A Taylor expansion about $\xi=0$ reveals how the entanglement dimensionality grows in the weak-squeezing regime.
Because $\xi=0$ is a stationary minimum of the Schmidt number, its first derivative vanishes there, making the entanglement dimensionality first-order insensitive to small deviations from the matching condition.
The leading variation is therefore quadratic in the squeezing parameter.
Expanding the relative Schmidt number up to fourth order and defining the fourth-order coefficient $\alpha_{N_+}^{N_-}$, we obtain
\begin{align} \label{eq:K(xi)-xi^4;sch-num}
  \dfrac{K_{N_+}^{N_-}(\xi)}{K_{N_+}^{N_-}(0)}
  &= 1+4\xi^2+\alpha_{N_+}^{N_-}\xi^4+O(\xi^6),\\
  \alpha_{N_+}^{N_-}
  &\define \dfrac{16}{3}
  - \dfrac{N_+ N_-}{(2N_+-1)(2N_--1)},
\end{align}
where $O(\xi^6)$ denotes a term bounded above by $\xi^6$.
The quadratic term, $4\xi^2$, is independent of the pump-mode indices, indicating a universal leading-order relative enhancement of the entanglement dimensionality.
The absolute increase is instead proportional to the mode-conserving Schmidt number $K_{N_+}^{N_-}(0)$ and therefore becomes larger for higher-order pump modes, as shown in Eq.~\eqref{eq:K-xi=0_approx;sch-num}.
The first explicit dependence on the pump-mode structure appears in the fourth-order coefficient $\alpha_{N_+}^{N_-}$.
For $p=0$, it takes its maximum value,
\begin{align}
\alpha_{N_+}^{N_-}=\frac{16}{3},
\qquad p=\min(N_+,N_-)=0.
\end{align}
For $p\geq1$, the coefficient increases monotonically with either mode index and satisfies
\begin{align}
\frac{13}{3}=\alpha_{1}^{1}
\leq \alpha_{N_+}^{N_-}
< \frac{61}{12},
\qquad N_+,N_-\geq1,
\end{align}
where the lower bound is attained at $N_+=N_-=1$, while the upper bound is approached asymptotically as both mode indices become large.

In the opposite limit of strong spatial squeezing, $\abs{\xi}\gg1$, the asymptotic expression derived in Appendix~\ref{app:schmidt} gives
\begin{widetext}
\begin{align} \label{eq:Ka-xi;sch-num}
  K_{N_+}^{N_-} (\xi)
  \approx \dfrac{1}{4} \ee^{4\abs{\xi}} \times
  \begin{cases}
    \frac{\pi}{2} \sqrt{\abs{l} + 1/4},
    & (p=0), \\
    \frac{\pi^2}{2}
    \frac{p+1/2}{\ln(p+3/4)+\gamma\subt{E}+4\ln(2)},
    & (p \geq 1, l = 0), \\
    \pi^2
    \frac{\sqrt{(p+1/4)(p+\abs{l}+1/4)}}
    {\ln\mleft[(p+3/4)(p+\abs{l}+1/4)/\abs{l}\mright]
    + \gamma\subt{E} + 6\ln(2)},
    & (p \geq 1, l \neq 0),
  \end{cases}
\end{align}
\end{widetext}
where $\gamma\subt{E}\approx0.577$ is the Euler--Mascheroni constant.
Equation~\eqref{eq:Ka-xi;sch-num} shows that all pump modes share the same exponential scaling, $K\propto\ee^{4\abs{\xi}}$, while the pump-mode dependence is contained in the prefactor.
For $p=0$, the prefactor scales as $\sqrt{\abs{l}}$ for large $\abs{l}$, whereas for large $p$ and fixed $l$ it grows approximately as $p/\ln p$.
Thus, radial and azimuthal pump excitations contribute differently to the entanglement dimensionality in the strong-squeezing regime.
This behavior contrasts with the correlation-width matching condition $\xi=0$, for which Eq.~\eqref{eq:K-xi=0_approx;sch-num} gives an approximately linear growth with $p$ at large radial order.
Spatial squeezing therefore reduces the radial-order dependence from approximately $p$ to $p/\ln p$, whereas the square-root dependence on $\abs{l}$ is retained asymptotically.

\begin{figure}
  \centering
  \includegraphics[width=.4\textwidth]{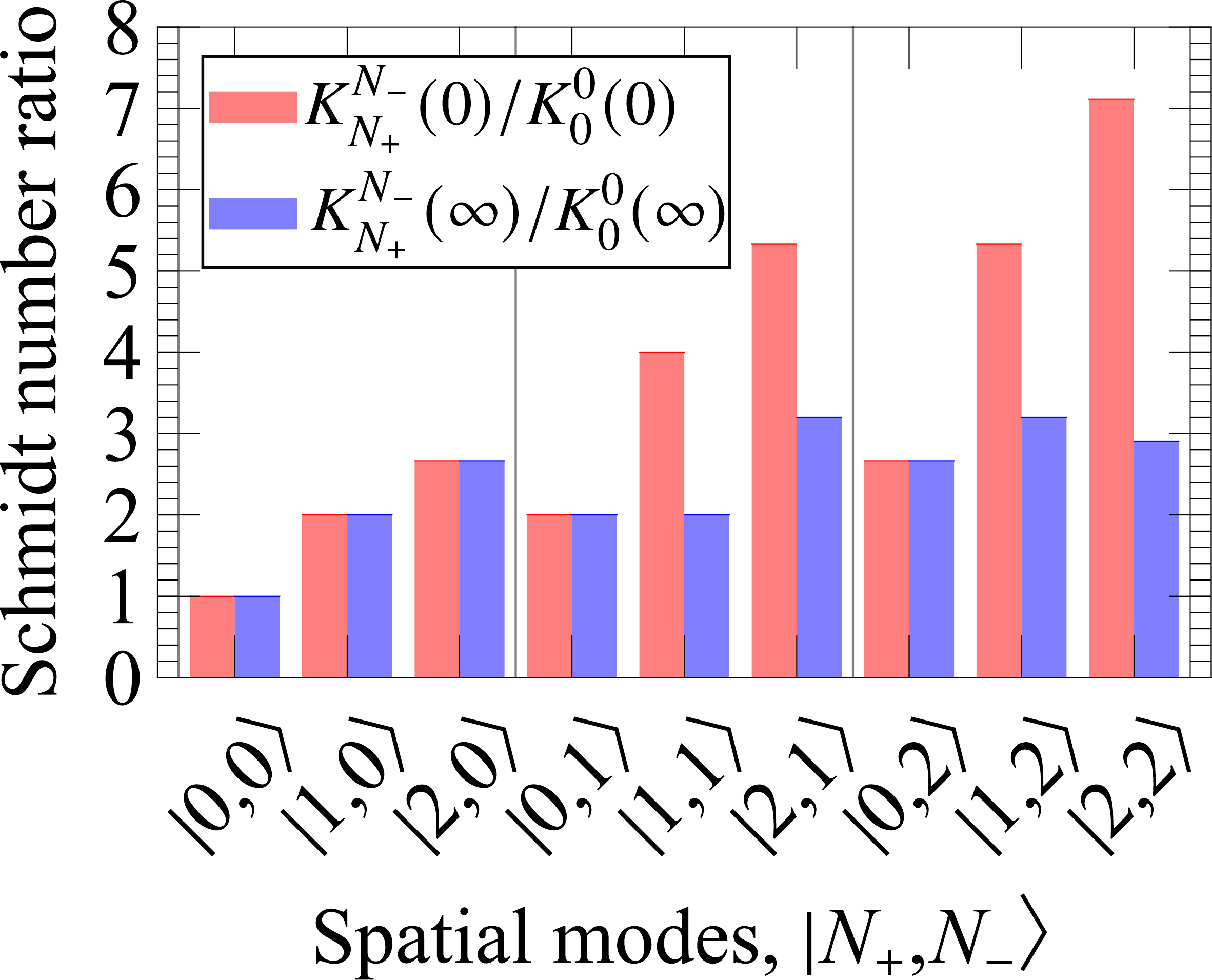}
  \caption{
    Comparison of the pump-mode-dependent Schmidt-number ratio
    between the correlation-width matching condition $\xi=0$ (red bars)
    and the strong-squeezing limit $\abs{\xi}\to\infty$ (blue bars).
    For $p=\min(N_+,N-)=0$, the ratio is independent of $\xi$ and is therefore preserved under spatial squeezing.
    For $p\geq 1$, the ratio decreases in the strong-squeezing limit, with the reduction being most pronounced for purely radial pump modes, $N_+=N_-=p$.
  }
  \label{fig:N=0-10_K;sch-num}
\end{figure}

Figure~\ref{fig:N=0-10_K;sch-num} compares the pump-mode-dependent Schmidt-number ratio $K_{N_+}^{N_-}(\xi)/K_0^0(\xi)$ between the correlation-width matching condition $\xi=0$ and the strong-squeezing limit $\abs{\xi}\to\infty$.
For $p=\min(N_+,N_-)=0$, only the $n=0$ term contributes in Eq.~\eqref{eq:K-xi;sch-num}, making the ratio independent of $\xi$; hence, the pump-mode-dependent enhancement is exactly preserved under spatial squeezing.
For $p\geq1$, in contrast, the ratio decreases in the strong-squeezing limit, with the reduction being most pronounced for purely radial pump modes.

\section{\label{sec:conc}Conclusion}

In conclusion, we have developed an algebraic framework for characterizing the spatial-mode entanglement generated by SPDC pumped with arbitrary higher-order LG modes.
By choosing the biphoton LG-basis radius as the geometric mean of the pump-beam radius and the crystal-induced correlation width, the spatial transformation of the biphoton state can be factorized into a spatial-mode beam-splitter transformation followed by spatial two-mode squeezing.
This decomposition separates the redistribution of the pump-mode excitations between the signal and idler from the spatial squeezing induced by the mismatch between the pump-beam radius and the correlation width.

Under the correlation-width matching condition, $\xi=0$, spatial squeezing vanishes and the two circular-mode numbers of the pump are independently conserved across the signal-idler pair.
As a consequence, both the OAM and the total spatial mode number are conserved, and the biphoton state is confined to a finite LG-mode manifold.
Under this condition, the LG-mode expansion also constitutes the Schmidt decomposition even for higher-order LG pumps, providing a natural modal basis for the spatial entanglement.
For a fundamental Gaussian pump, this condition yields a separable biphoton state, whereas a higher-order LG pump remains spatially entangled through the coherent redistribution of the pump-mode excitations by the spatial-mode beam splitter.
Away from the matching condition, spatial squeezing removes the fixed-total-mode-number constraint while preserving the mode-difference invariants and, consequently, OAM conservation.

We have further derived an analytical expression for the Schmidt number for arbitrary LG pump modes and squeezing strengths.
At the matching condition, the Schmidt number is determined entirely by the pump-mode structure and exhibits distinct asymptotic dependences on the radial and azimuthal indices.
In the weak-squeezing regime, the leading relative increase is universal and quadratic in the squeezing parameter, while the first explicit pump-mode dependence appears at fourth order.
In the strong-squeezing regime, all pump modes share the universal exponential scaling $K\propto\exp(4|\xi|)$, whereas the pump-mode dependence is contained in the prefactor.
In particular, the radial-order dependence changes from approximately $p$ at the matching condition to $p/\ln p$ under strong squeezing, while the asymptotic square-root dependence on $|l|$ is retained.

These results establish an algebraic route from the spatial structure of the pump to the conservation laws and entanglement dimensionality of the generated biphoton state.
The framework provides a basis for systematically engineering and analyzing high-dimensional spatial entanglement in structured-light-pumped SPDC.

\appendix
\section{\label{app:squeezing}Derivation of the Spatial Mode Beam Splitter and Squeezing Transformations}

\begin{figure*}
  \centering
  \includegraphics[width=1.\textwidth]{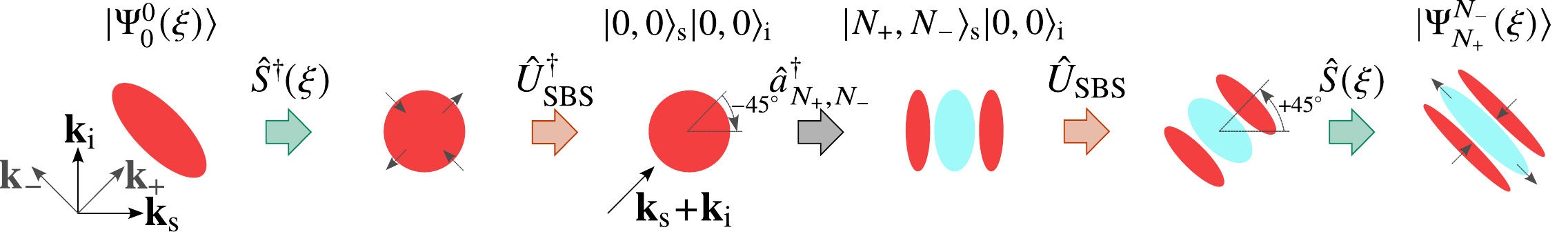}
  \caption{
    The SPDC biphoton pumped by the fundamental Gaussian mode $\ket{\Psi_0^0(\xi)}$ is disentangled using the process of $\hat{S}^\dagger(\xi)$ and $\hat{U}\subt{SBS}^\dagger$.
    Subsequently, after applying the raising operator $\hat{a}^\dagger_{N_+,N_-}$, the forward process $\hat{S}(\xi)$ and $\hat{U}\subt{SBS}$ are applied to generate the higher-order biphoton state $\ket{\Psi_{N_+}^{N_-}(\xi)}$.
  }
  \label{fig:SPDC-process;squeezing}
\end{figure*}

The operator process for generating the biphoton state $\ket{\Psi_{N_+}^{N_-}(\xi)}$ via SPDC pumped by a monochromatic pure LG mode is illustrated in Fig.~\ref{fig:SPDC-process;squeezing}. This formulation reduces the calculation of the biphoton state generated by a higher-order LG mode, which conventionally requires a complicated overlap integral, to an algebraic problem involving a raising operator. First, the biphoton wavefunction in momentum space for the state $\ket{\Psi_0^0(\xi)}$, generated by SPDC pumped by a monochromatic fundamental Gaussian mode, exhibits a strong peak when both the momentum-conservation law between the pump field and the two photons, $\mathbf{k}\subt{p} = \mathbf{k}\subt{s} + \mathbf{k}\subt{i}$, and the phase-matching condition, $\mathbf{k}\subt{s} - \mathbf{k}\subt{i} \approx 0$, are satisfied.

Letting the $1/\ee^2$ widths determined by the momentum conservation law and the phase-matching condition be $2w\subt{p}^{-1}$ and $2w_-^{-1}$, respectively, this state can be interpreted as the following geometric scale transformation. That is, it is equivalent to the state obtained by applying spatial squeezing to the uncorrelated biphoton state $\ket{0,0}\subt{s} \ket{0,0}\subt{i}$ corresponding to the beam radius $w_0$ in Eq.~\eqref{eq:w0;sbs} defined by the geometric mean, thereby expanding the distribution width along the sum-coordinate $\mathbf{k}_+ \define (\mathbf{k}\subt{s} + \mathbf{k}\subt{i})/\sqrt{2}$ axis by a factor of $\ee^\xi$ and contracting the distribution width along the difference-coordinate $\mathbf{k}_- \define (\mathbf{k}\subt{s} - \mathbf{k}\subt{i})/\sqrt{2}$ axis by a factor of $\ee^{-\xi}$. Here, the sum and difference coordinate system $(\mathbf{k}_+, \mathbf{k}_-)$ is tilted by $+45^\circ$ in wavenumber space with respect to the coordinate system of the individual photons $(\mathbf{k}\subt{s},\mathbf{k}\subt{i})$. Therefore, by rotating the $\mathbf{k}\subt{s}$-$\mathbf{k}\subt{i}$ plane by $-45^\circ$, the alteration of the spatial distribution of the pump field $E\subt{p}(\mathbf{k}\subt{p})$ can be treated independently simply as a change in the distribution along the $\mathbf{k}\subt{s}$ axis. 

Based on the above geometric picture, the process of inversely transforming the correlated biphoton state $\ket{\Psi_0^0(\xi)}$ into the more tractable uncorrelated state $\ket{0,0}\subt{s} \ket{0,0}\subt{i}$ can be described as follows. It suffices to apply the two-mode squeezing operator $\hat{S}^\dagger(\xi)$ that preserves OAM conservation to disentangle the quantum entanglement, and rotate the distribution by $-45^\circ$ using the SBS operator $\hat{U}\subt{SBS}^\dagger$ to align the axis of the momentum conservation law $\mathbf{k}_+ \propto \mathbf{k}\subt{p}$ to the direction of the wavenumber basis $\mathbf{k}\subt{s}$ (Fig.~\ref{fig:SPDC-process;squeezing}). 

Using this simplified spatial mode, when a mode transformation is performed from the fundamental Gaussian mode to the pump field $E\subt{p}(\mathbf{k}\subt{p}) = \text{LG}_{N_+}^{N_-}(\mathbf{k}\subt{p})$ whose spatial mode has been raised by the raising operator $\hat{a}_{N_+,N_-}^\dagger$, the biphoton state $\ket{0,0}\subt{s} \ket{0,0}\subt{i}$ becomes $\ket{N_+,N_-}\subt{s} \ket{0,0}\subt{i}$. From here, to return to the target SPDC biphoton state $\ket{\Psi_{N_+}^{N_-}(\xi)}$, one simply needs to apply the inverse process $\hat{S}(\xi)\hat{U}\subt{SBS}$ (Fig.~\ref{fig:SPDC-process;squeezing}). 

The mode raising via the LG mode raising operator $\hat{a}_{N_+,N_-}^\dagger$ [Eq.~\eqref{eq:LGmode;main}] demonstrated in the above process is not unique to LG modes. Because the ladder operators of a two-dimensional harmonic oscillator are interconnected by $\mathrm{SU}(2)$ transformations, and each point on the $\mathrm{SU}(2)$ Poincar{\'e} sphere corresponds to Hermite--Gaussian (HG) modes, LG modes, and general Gaussian modes that continuously interpolate between them, the mode-raising structure obtained here naturally extends from LG modes to arbitrary Gaussian modes as a whole~\cite{Padgett:99,Agarwal:99,Habraken:10,10.1098/rsta.2015.0441,PhysRevA.102.031501}. In particular, when the squeezing parameter is $\xi=0$, the mode selection rule~\eqref{eq:N+,N-;LGmode_decom} applies. This selection rule need not be confined to raising operators for specific LG modes; rather, by employing higher-order Gaussian modes unified within the $\mathrm{SU}(2)$ framework, it can be understood as a universal mode-raising rule.
\section{\label{app:algebraic}Algebraic Derivation of the SPDC State using the Spatial Mode Difference Operator}

The relative mode-difference operators $\hat{\Delta}$ and $\hat{\Delta}'$, corresponding to the mode-number differences $\Delta \hat{N}_{\text{s}+,\text{i}-}$ and $\Delta \hat{N}_{\text{s}-,\text{i}+}$, respectively, commute with the squeezing operator $\hat{S}(\xi)$ according to Eq.~\eqref{eq:[S,N_Delta];LGmode_decom}.
Based on Eq.~\eqref{eq:UBSpsi;LGmode_decom}, we re-express the summation over $(m,n)$ to separate the state into eigenstates in terms of $m$ and the relative mode difference $\Delta$:
\begin{align} \label{eq:Psi0-decomp;alg_spdc}
  \ket{\Psi_{N_+}^{N_-}(0)}
  &= \sum_{\Delta=-N_-}^{+N_+}
  \sqrt{Z_\Delta} \ket{\Phi_\Delta(0)}
\end{align}
Here, $Z_\Delta$ is the coefficient satisfying $\braket{\Phi_{\Delta_1} (0) \vert{} \Phi_{\Delta_2} (0)} = \delta_{\Delta_1}^{\Delta_2}$, which corresponds to the binomial probability amplitude for which a specific modal difference $\Delta$ is obtained upon spatial decomposition.
From Vandermonde's identity, the normalization coefficient is given by
\begin{align}
  Z_\Delta
  &= \sum_{m=\max(0,\Delta)}^{\min(N_+,N_-+\Delta)}
    \dfrac{1}{2^N}
    \binom{N_+}{m} \binom{N_-}{N_--m+\Delta}
    \nonumber \\
  &= \dfrac{1}{2^N} \binom{N}{N_-+\Delta}.
\end{align}
where $\delta_m^n$ is the Kronecker delta.
The orthogonal substates are defined to satisfy the eigenvalue equations
\begin{align}
  \hat{\Delta}\ket{\Phi_\Delta(0)}
  &= \Delta\ket{\Phi_\Delta(0)}, \\
  \hat{\Delta}'\ket{\Phi_\Delta(0)}
  &= (\Delta-l)\ket{\Phi_\Delta(0)},
\end{align}
such that their orthogonality is maintained even after applying the squeezing operator $\hat{S}(\xi)$:
\begin{align}
  \ket{\Phi_\Delta (0)}
  &= \sum_{m=\max(0,\Delta)}^{\min(N_+,N_-+\Delta)}
    \sqrt{\lambda_{m}^{(\Delta)}}
    \ket{m}_{\text{s}+} \ket{m\!-\!\Delta}_{\text{i}-}
    \nonumber \\ & \quad \otimes
    \ket{N_-\!-\!m\!+\!\Delta}_{\text{s}-}
    \ket{N_+\!-\!m}_{\text{i}+}.
\end{align}
The normalized binomial weight is given by:
\begin{align}
  \lambda_{m}^{(\Delta)}
  \define \binom{N_+}{m} \binom{N_-}{m-\Delta} \binom{N}{N_-+\Delta}^{-1}.
\end{align}

In the presence of spatial squeezing $\xi \neq 0$, the general SPDC state is obtained by applying the two-mode squeezing operator $\hat{S}(\xi)$ to the decomposed sub-states:
\begin{align} \label{eq:Phi-xi;alg_spdc}
  \ket{\Phi_\Delta (\xi)} &= \hat{S}(\xi) \ket{\Phi_\Delta (0)}.
\end{align}
Because $\hat{\Delta}$ commutes with $\hat{S}(\xi)$, the eigenvalue $\Delta$ remains constant during this transition. By expanding the squeezing operator using its known matrix elements $S_{a_1,a_2}^{b_1,b_2} = \braket{b_1,b_2|\hat{S}(\xi)|a_1,a_2}$ in the Fock basis~\cite{gu2026analyticalfockrepresentationtwomode}, we obtain the algebraic form of the generalized biphoton state:
\begin{align} \label{eq:Phi_Delta_xi;alg_spdc}
  \ket{\Phi_\Delta (\xi)}
  &= \sum_{m}
    \sqrt{\lambda_m^{(\Delta)}}
    \ket{\Phi_{m,\Delta}^{(+)}(\xi)}
    \ket{\Phi_{m,\Delta}^{(-)}(\xi)} \\
    \label{eq:Phi_Delta_xi_ab;alg_spdc}
  &= \sum_{a=\max(0,\Delta)}^{\infty}
    \sum_{b=\max(0,\Delta-l)}^{\infty}
    C_{a,b}^{(\Delta)}(\xi)
    \nonumber \\ &\quad \times
    \ket{a,b}_{\text{s}}
    \ket{a-\Delta,b-\Delta+l}_{\text{i}},
\end{align}
where the coefficient $C_{a,b}^{(\Delta)}(\xi)$ is a probability amplitude dependent on $\xi$, $m$ acts as an internal DoF, and the partial states $\ket{\Phi_{m,\Delta}^{(+)}(\xi)}$ and $\ket{\Phi_{m,\Delta}^{(-)}(\xi)}$ are
\begin{align} \label{eq:Phi-Op-mDelta-plus;alg_spdc}
  \ket{\Phi_{m,\Delta}^{(+)}(\xi)}
  &\define \hat{S}_+(\xi) \ket{m}_{\text{s}+}\ket{m-\Delta}_{\text{i}-}
    \\ \label{eq:Phi-mDelta-plus;alg_spdc}
  &= \sum_{a}
    S_{m,m-\Delta}^{a,a-\Delta} \ket{a}_{\text{s}+}\ket{a-\Delta}_{\text{i}-},
    \\ \label{eq:Phi-Op-mDelta-minus;alg_spdc}
  \ket{\Phi_{m,\Delta}^{(-)}(\xi)}
  &\define \hat{S}_-(\xi)
    \ket{N_-\!-\!m\!+\!\Delta}_{\text{s}-}\ket{N_+\!-\!m}_{\text{i}+}
    \\ \label{eq:Phi-mDelta-minus;alg_spdc}
  &= \sum_{b}
    S_{N_--m+\Delta,N_+-m}^{b,b-\Delta+l} \ket{b}_{\text{s}-}\ket{b\!-\!\Delta\!+\!l}_{\text{i}+} .
\end{align}
Here the two-mode squeezing operators are given by
\begin{align}
  \hat{S}_\pm(\xi)
  &\define \ee^{\xi(\hat{a}_{\text{s}\pm}^\dagger \hat{a}_{\text{i}\mp}^\dagger - \text{H.c.})}.
\end{align}
These substates in Eqs.~\eqref{eq:Phi-Op-mDelta-plus;alg_spdc} and \eqref{eq:Phi-Op-mDelta-minus;alg_spdc} possess the important property of being orthogonal,
\begin{align} \label{eq:Phi-Op-Orth;alg_spdc}
  \braket{\Phi_{m_1,\Delta_1}^{(\pm)}(\xi)|\Phi_{m_2,\Delta_2}^{(\pm)}(\xi)}
  = \delta_{m_1}^{m_2}
    \delta_{\Delta_1}^{\Delta_2},
\end{align}
when the indices $(m,\Delta)$ differ.

From Eq.~\eqref{eq:Phi_Delta_xi;alg_spdc}, the total spatial mode difference $N_\Delta$ evaluates to $2\Delta - l$. Because this difference is conserved under $\hat{S}(\xi)$ and the initial state $\ket{\Phi_\Delta(0)}$ obeys the sum constraint $N\subt{s} + N\subt{i} = N$ with non-negative mode numbers $N\subt{s}, N\subt{i} \ge 0$, it follows that $N_\Delta$ is fundamentally restricted to the finite range
\begin{align} \label{eq:N_Delta;alg_spdc}
  -N \le N_\Delta \le N.
\end{align}
Here, the expansion coefficient is determined by the discrete convolution of the unsqueezed distribution with the squeezing matrix elements:
\begin{align}
  C_{a,b}^{(\Delta)} (\xi)
  &= \sum_{m}
    \sqrt{\lambda_{m}^{(\Delta)}}
    S_{m,m-\Delta}^{a,a-\Delta}
    S_{N_--m+\Delta,N_+-m}^{b,b-\Delta+l}.
\end{align}
From Eqs.~\eqref{eq:Psi0-decomp;alg_spdc}, \eqref{eq:Phi-xi;alg_spdc}, and \eqref{eq:Phi_Delta_xi_ab;alg_spdc}, the independent discrete modes are indexed by the set $(a,b,\Delta)$, and each probability amplitude $C_{a,b}^{(\Delta)} (\xi)$ is expressed as a finite sum over $m$.

This algebraic formulation of Eq.~\eqref{eq:Phi_Delta_xi_ab;alg_spdc} provides a mathematically precise description of the continuous-variable entanglement generated in SPDC, relying solely on the Fock basis and invariant mode differences rather than integrated Gaussian field profiles.

\section{\label{app:schmidt}Derivation of the Schmidt Number via Generating Functions}

In the context of SPDC, the spatial entanglement between the signal and idler photons can be quantified by the Schmidt number $K$. Because the biphoton wavefunction $\psi_{N_+}^{N_-}(\bold{k}\subt{s}, \bold{k}\subt{i})$ in our formalism is not necessarily normalized, the exact Schmidt number is given by the ratio of the squared trace of the unnormalized density matrix to its purity:
\begin{align} \label{eq:K def;schmidt}
  K \define \frac{(\text{Tr}[\rho])^2}{\text{Tr}[{\rho\subt{s}}^2]}.
\end{align}
The continuous unnormalized trace and spatial purity are expressed via the overlap integrals of the biphoton momentum state amplitudes:
\begin{align}
  \text{Tr}[\rho]
  &\!=\! \int\! \dfrac{\odif{\bold{k}\subt{s}}}{(2\pi)^2} \int\! \dfrac{\odif{\bold{k}\subt{i}}}{(2\pi)^2} \psi_{N_+}^{N_-}(\bold{k}\subt{s}, \bold{k}\subt{i}) \psi_{N_+}^{N_-}(\bold{k}\subt{s}, \bold{k}\subt{i})^*, \\
  \text{Tr}[{\rho\subt{s}}^2]
  &\!=\! \int\! \dfrac{\odif{\bold{k}\subt{s}}}{(2\pi)^2} \int\! \dfrac{\odif{\bold{k}'\subt{s}}}{(2\pi)^2} \int\! \dfrac{\odif{\bold{k}\subt{i}}}{(2\pi)^2} \int\! \dfrac{\odif{\bold{k}'\subt{i}}}{(2\pi)^2}
  \nonumber \\ & \, \times \!
  \psi_{N_+}^{N_-}\!(\bold{k}\subt{s}, \bold{k}\subt{i}) \psi_{N_+}^{N_-}\!(\bold{k}'\subt{s}, \bold{k}\subt{i})^*
  \psi_{N_+}^{N_-}\!(\bold{k}'\subt{s}, \bold{k}'\subt{i}) \psi_{N_+}^{N_-}\!(\bold{k}\subt{s}, \bold{k}'\subt{i})^*,
\end{align}
where the complex conjugate is $(\cdot)^*$.

To analytically derive these quantities for an SPDC state pumped by a generalized LG mode, we employ the coherent state generating function. Specifically, we substitute the pump field LG modes in the biphoton amplitudes $\psi_{N_+}^{N_-}$ with the generating function $G_0$:
\begin{align}
  G_0(\alpha, \beta; \bold{k})
  &\define \sqrt{\pi} \sum_{N_+=0}^\infty \sum_{N_-=0}^\infty \frac{\alpha^{N_+} \beta^{N_-}}{\sqrt{N_+! N_-!}} \text{LG}_{N_+}^{N_-}(\bold{k}) \nonumber \\
  &= \ee^{
    \sqrt{2}\alpha \bold{e}\subt{L} \cdot \bold{k} + \sqrt{2}\beta \bold{e}\subt{R} \cdot \bold{k} - \alpha \beta
    }
    \ee^{-\norm{\bold{k}}^2/2},
\end{align}
where transverse wavevector $\bold{k}$ is normalized by $\sqrt{2}{w\subt{p}}^{-1}$ in this section, the inner product of two vectors $\bold{a}=(a_x,a_y)^\top$ and $\bold{b}=(b_x,b_y)^\top$ is given by $\bold{a}\cdot\bold{b} = {a_x} b_x + {a_y} b_y$, $\bold{e}\subt{L,R}\!=\!(\bold{e}\subt{H}\pm\ii \bold{e}\subt{V})/\sqrt{2}$ are the left- and right-handed circular polarization unit vectors, and $\bold{e}\subt{H,V}$ are the horizontal and vertical polarization unit vectors.
Because $\text{Tr}[\rho]$ contains two amplitudes and $\text{Tr}[{\rho\subt{s}}^2]$ contains four, we introduce independent generating parameters for each amplitude: $(\alpha_1, \beta_1)$ and $(\alpha_2, \beta_2)$ for $\text{Tr}[\rho]$, and extending up to $(\alpha_4, \beta_4)$ for $\text{Tr}[{\rho\subt{s}}^2]$. To simplify the calculation of partial derivatives, we treat these complex generating parameters as purely real variables.

By arranging these parameters into vectors $\bold{t} = (\alpha_1, \alpha_2, \beta_1, \beta_2)^\top$ and $\bold{T} = (\alpha_1, \alpha_2, \alpha_3, \alpha_4, \beta_1, \beta_2, \beta_3, \beta_4)^\top$, the generating function transforms the integrands into purely Gaussian exponents, allowing the four- and eight-dimensional momentum integrals to be evaluated analytically:
\begin{widetext}
\begin{align}
  \text{Tr}[\rho(\bold{t})]
  &= \int\! \dfrac{\odif{\bold{k}\subt{s}}}{(2\pi)^2} \int\! \dfrac{\odif{\bold{k}\subt{i}}}{(2\pi)^2}
    G_0(\alpha_1,\beta_1; \bold{k}\subt{s}\!+\!\bold{k}\subt{i})
    \ee^{-\gamma^{-2} \norm{\bold{k}\subt{s}-\bold{k}\subt{i}}^2/2}
    G_0(\alpha_2,\beta_2; \bold{k}\subt{s}\!+\!\bold{k}\subt{i})^*
    \ee^{-\gamma^{-2} \norm{\bold{k}\subt{s}-\bold{k}\subt{i}}^2/2} \nonumber \\
  &= \frac{\gamma^2}{16\pi(2\pi)^2} \exp\mleft[ -\frac{1}{2}\bold{t}^\top \bold{C} \bold{t} \mright], \\
  \text{Tr}[{\rho\subt{s}}^2(\bold{T})]
  &= \int\! \dfrac{\odif{\bold{k}\subt{s}}}{(2\pi)^2} \int\! \dfrac{\odif{\bold{k}\subt{i}}}{(2\pi)^2} \int\! \dfrac{\odif{\bold{k}\subt{s}'}}{(2\pi)^2} \int\! \dfrac{\odif{\bold{k}\subt{i}'}}{(2\pi)^2}
    G_0(\alpha_1,\beta_1; \bold{k}\subt{s}\!+\!\bold{k}\subt{i})
    \ee^{-\gamma^{-2} \norm{\bold{k}\subt{s}-\bold{k}\subt{i}}^2/2}
    G_0(\alpha_2,\beta_2; \bold{k}\subt{s}'\!+\!\bold{k}\subt{i})^*
    \ee^{-\gamma^{-2} \norm{\bold{k}\subt{s}'-\bold{k}\subt{i}}^2/2} \nonumber \\ & \quad \times
    G_0(\alpha_3,\beta_3; \bold{k}\subt{s}'\!+\!\bold{k}\subt{i}')
    \ee^{-\gamma^{-2} \norm{\bold{k}\subt{s}'-\bold{k}\subt{i}'}^2/2}
    G_0(\alpha_4,\beta_4; \bold{k}\subt{s}\!+\!\bold{k}\subt{i}')^*
    \ee^{-\gamma^{-2} \norm{\bold{k}\subt{s}-\bold{k}\subt{i}'}^2/2} \nonumber \\
  &= \dfrac{1}{K_0^0(\xi)} \mleft[ \frac{\gamma^2}{16\pi(2\pi)^2} \mright]^2
    \exp\mleft[ -\frac{1}{2}\bold{T}^\top \bold{D} \bold{T} \mright].
\end{align}
\end{widetext}
Here the squeezing parameter is $\gamma \define \ee^{2\xi}$. The coupling matrices $\bold{C}$ and $\bold{D}$ dictate the cross-correlations generated by the phase-matching condition and spatial squeezing. Using the $2\times 2$ identity matrix $\hat{\bold{I}}$ and the Pauli-X matrix $\hat{\bold{X}}$, they are explicitly given by:
\begin{align}
  \bold{C}
  &\!=\! - \hat{\bold{I}} \!\otimes\! \hat{\bold{X}}, \\
  \bold{D}
  &\!=\! - \frac{1}{2} \mleft[
    \hat{\bold{I}} \!\otimes\! (\hat{\bold{I}} \!+\! \hat{\bold{X}}) \!\otimes\! \hat{\bold{X}}
    \!-\! \frac{1-\gamma^2}{1+\gamma^2} \hat{\bold{X}} \!\otimes\! (\hat{\bold{I}} \!-\! \hat{\bold{X}}) \!\otimes\! \hat{\bold{I}}
    \mright].
\end{align}

The exponential part of the spatial purity $\Tr[\rho_s^2]$ can be simplified by applying the transform to the generating parameters, which diagonalizes the coupling matrix $\bold{D}$:
\begin{align}
  \bold{D}'
  &\define \hat{U} \bold{D} \hat{U}^\top \nonumber \\ \label{eq:Dd;schmidt}
  &= - \mleft[
      \hat{\bold{I}} \!\otimes\! \hat{P}_0 \!\otimes\! \hat{\bold{X}}
      + \mu_2 \hat{\bold{X}} \!\otimes\! \hat{P}_1 \!\otimes\! \hat{\bold{I}}
    \mright],
\end{align}
where $\hat{P}_0 \define \text{diag}(1,0)$ and $\hat{P}_1 \define \text{diag}(0,1)$ are projection matrices, and $\hat{\bold{Z}}$ is the Pauli-Z matrix, $\hat{U} \define \hat{\bold{I}}\otimes\hat{\bold{H}}\otimes\hat{\bold{I}}$ is the equivalent action of the Hadamard transform on the second qubit. The transformed generating parameters are defined as
\begin{align} \label{eq:Td;schmidt}
  \bold{T}'
  \define \hat{U} \bold{T},
\end{align}
and the nabla operator of the parameter vector $\bold{T}$ can be rewritten in terms of the transformed parameters $\bold{T}'$ as
\begin{align} \label{eq:nabla trans;schmidt}
  \bold{\nabla}_{\bold{T}}
  &= \hat{U} \bold{\nabla}_{\bold{T}'}.
\end{align}
Thus, we can simplify the products of the derivatives
\begin{align}
  \prod_{i=1}^4 \dfrac{\partial}{\partial\alpha_i}
  &= \dfrac{1}{4} \mleft[
      \dfrac{\partial^2}{\partial{\alpha'_1}^2}
      - \dfrac{\partial^2}{\partial{\alpha'_3}^2}
    \mright]
    \mleft[
      \dfrac{\partial^2}{\partial{\alpha'_2}^2}
      - \dfrac{\partial^2}{\partial{\alpha'_4}^2}
    \mright], \\
  \prod_{i=1}^4 \dfrac{\partial}{\partial\beta_i}
  &= \dfrac{1}{4} \mleft[
      \dfrac{\partial^2}{\partial{\beta'_1}^2}
      - \dfrac{\partial^2}{\partial{\beta'_3}^2}
    \mright]
    \mleft[
      \dfrac{\partial^2}{\partial{\beta'_2}^2}
      - \dfrac{\partial^2}{\partial{\beta'_4}^2}
    \mright].
\end{align}

Therefore, the exponential part of the spatial purity can be expressed as a product of four independent exponential functions:
\begin{align}
  - \dfrac{1}{2} \bold{T}^\top \bold{D} \bold{T}
  &= - \dfrac{1}{2} \bold{T}'^\top \bold{D}' \bold{T}' \nonumber \\
  &= (\alpha'_1 \alpha'_2 + \beta'_1 \beta'_2)
    + \mu_2 (\alpha'_3 \beta'_3 + \alpha'_4 \beta'_4).
\end{align}

The physical trace and purity corresponding to the specific LG mode $\ket{N_+,N_-}$ are then extracted by applying partial derivatives to the generating traces evaluated at the origin $\bold{t}=0$ and $\bold{T}=0$:
\begin{align}
  \text{Tr}[\rho]
  &= \frac{1}{\pi N_+! N_-!} \mleft[ \prod_{i=1}^2 \frac{\partial^{N_+}}{\partial \alpha_i^{N_+}} \frac{\partial^{N_-}}{\partial \beta_i^{N_-}} \text{Tr}[\rho(\bold{t})] \mright]_{\bold{t}=0}, \label{eq:Tr_rho;schmidt} \\
  \text{Tr}[{\rho\subt{s}}^2]
  &= \frac{1}{(\pi N_+! N_-!)^2} \mleft[ \prod_{i=1}^4 \frac{\partial^{N_+}}{\partial \alpha_i^{N_+}} \frac{\partial^{N_-}}{\partial \beta_i^{N_-}} \text{Tr}[{\rho\subt{s}}^2(\bold{T})] \mright]_{\bold{T}=0}. \label{eq:Tr_rho_s2;schmidt}
\end{align}
The exponential part of the trace of the unnormalized density matrix $\Tr[\rho]$ can be expanded as a power series in the generating parameters, which allows the extraction of the specific LG mode contributions:
\begin{align} \label{eq:exptCt;schmidt}
  \prod_{i=1}^4 \mleft.
    \dfrac{\partial^N \exp\mleft[ - \dfrac{1}{2} \bold{t}^\top\bold{C}\bold{t} \mright]}{\partial\alpha_i^{N_+}\partial\beta_i^{N_-}} \mright|_{\bold{t}=0}
    &= N_+! N_-!.
\end{align}

\begin{widetext}
\begin{align}
  \prod_{i=1}^4 \mleft.
    \dfrac{\partial^N \exp\mleft[ - \dfrac{1}{2} \bold{T}^\top\bold{D}\bold{T} \mright]}{\partial\alpha_i^{N_+}\partial\beta_i^{N_-}} \mright|_{\bold{T}=0}
  &= \dfrac{1}{4^N} \sum_{m=0}^{N_+} \sum_{m'=0}^{N_+} \sum_{n=0}^{N_-} \sum_{n'=0}^{N_-}
    \binom{N_+}{m}\binom{N_+}{m'}
    \binom{N_-}{n}\binom{N_-}{n'}
    (-1)^{m+n+m'+n'}
    \nonumber \\ & \quad \times
    (2N_+-2m)! \delta_{m}^{m'}
    (2N_--2n)! \delta_{n}^{n'}
    (2m)! \mu_2^{2m} \delta_{m}^{n}
    (2n)! \mu_2^{2n} \delta_{m'}^{n'}
    \nonumber \\ \label{eq:expTDT;schmidt}
  &= (N_+! N_-!)^2
    \sum_{n=0}^{\min(N_+,N_-)}
    \mleft[
      K_{N_+-n}^{N_--n}(0)
      K_{n}^{n}(0)
    \mright]^{-1}
    {\mu_2}^{4n}.
\end{align}
\end{widetext}

After substituting Eqs.~\eqref{eq:exptCt;schmidt} and \eqref{eq:expTDT;schmidt} into Eqs.~\eqref{eq:Tr_rho;schmidt} and \eqref{eq:Tr_rho_s2;schmidt}, respectively, and then inserting the resulting expressions into the definition of the Schmidt number, Eq.~\eqref{eq:K def;schmidt}, we obtain Eq.~\eqref{eq:K-xi;sch-num}:
\begin{align}
  K_{N_+}^{N_-} (\xi)
  &= K_0^0(\xi) \mleft(
    \sum_{n=0}^{p}
    \mleft[
      K_{N_+-n}^{N_--n}(0)
      K_{n}^{n}(0)
    \mright]^{-1}
    {\mu_2}^{4n}
  \mright)^{-1}.
\end{align}

By applying the Pochhammer symbol $\poc{x}{n} \define \Gamma(x+n)/\Gamma(x)$ with the Gamma function $\Gamma$, Eq.~\eqref{eq:K-xi=0;sch-num} can be rewritten as
\begin{align}
  K_{N_+}^{N_-}(0)
  = \dfrac{\poc{1}{N_+}}{\poc{(\frac{1}{2})}{N_+}}
    \dfrac{\poc{1}{N_-}}{\poc{(\frac{1}{2})}{N_-}}.
\end{align}
Eq.~\eqref{eq:K-xi;sch-num} with the Pochhammer function can be described by
\begin{align}
  \dfrac{1}{K_{N_+}^{N_-}(\xi)}
  &= \dfrac{1}{K_0^0(\xi)} \sum_{n=0}^{p}
    \dfrac{\poc{(\frac{1}{2})}{N_+-n}}{\poc{1}{N_+-n}}
    \dfrac{\poc{(\frac{1}{2})}{N_--n}}{\poc{1}{N_--n}}
    \dfrac{\poc{(\frac{1}{2})}{n}}{\poc{1}{n}}
    \dfrac{\poc{(\frac{1}{2})}{n}}{\poc{1}{n}}
    {\mu_2}^{4n} \nonumber \\
    \label{eq:K-xi temp;schmidt}
  &= \dfrac{[K_{N_+}^{N_-}(0)]^{-1}}{K_0^0(\xi)}
    \pFq{4}{3}{-N_+,-N_-,\frac{1}{2},\frac{1}{2}}{\frac{1}{2}-N_+,\frac{1}{2}-N_-,1}{{\mu_2}^4},
\end{align}
where the ${}_{p}F_{q}$ generalized hypergeometric function was used~\cite{Erdelyi1953}
\begin{align} \label{eq:pFq;schmidt}
  \pFq{p}{q}{a_1,a_2,\ldots,a_p}{b_q,b_2, \ldots, b_q}{z}
  \define \sum_{n=0}^{\infty}
    \dfrac{\poc{a_1}{n}\poc{a_2}{n} \cdots \poc{a_p}{n}}{\poc{b_1}{n}\poc{b_2}{n} \cdots \poc{b_q}{n}} \dfrac{z^n}{n!}.
\end{align}
Furthermore, the Hadamard product of two arbitrary functions $f(z) = \sum_{n=0}^{\infty} f_n z^n$ and $g(z) = \sum_{n=0}^{\infty} g_n z^n$ defined by
\begin{align} \label{eq:H prod;schmidt}
  (f \odot g)(z)
  \define \sum_{n=0}^{\infty} f_n g_n z^n,
\end{align}
Eq.~\eqref{eq:K-xi temp;schmidt} can be represented by
\begin{align}
  K_{N_+}^{N_-}(\xi)
  &= K_0^0(\xi)K_{N_+}^{N_-}(0)
    \nonumber \\ & \hspace{-1em} \times
    \mleft\{\pFq{2}{1}{-N_+,\frac{1}{2}}{\frac{1}{2}-N_+}{\mu_2^4} \!\odot\!
    \pFq{2}{1}{-N_-,\frac{1}{2}}{\frac{1}{2}-N_-}{\mu_2^4}\mright\}^{-1}.
\end{align}

\subsection{Approximate Schmidt number in the large pump mode and high squeezing limit}

In this subsection, we present the detailed derivation of the asymptotic behavior of the Schmidt number $K_{N_+}^{N_-}(\xi)$ in the limit of large pump modes, absent OAM, and high squeezing.

\subsubsection{Approximate Schmidt number in the large pump radial-mode and high squeezing limit}

We consider the regime of highly excited, OAM-less pure radial pump modes, defined by $p = N_+ = N_-$ and $l = 0$, under the strong squeezing limit $\xi \gg 1$. In this case, the combinatorial sum in the inverse Schmidt number ratio is evaluated by the convolution:
\begin{align} \label{eq:K-ratio_p;schmidt}
  A_p
  &= \dfrac{K_0^0(\xi)}{K_{N_+}^{N_-}(\xi)}
  \approx \sum_{n=0}^p {a_n}^2 {a_{p-n}}^2,
\end{align}
where the expansion coefficient $a_n$ is defined as
\begin{align}\label{eq:an;schmidt}
  a_n  \define \dfrac{1}{4^n} \binom{2n}{n}.
\end{align}
The Stirling expansion of the central binomial coefficient $a_n$ for large $n$ is given by~\cite{elezovic2014asymptotic}
\begin{align} \label{eq:an-asy;schmidt}
  a_n
  &= \dfrac{1}{\sqrt{\pi (n + 1/4)}} \mleft[
    1 - \dfrac{1}{64(n+1/4)^2} + O(n^{-4})
    \mright],
\end{align}
where we express the asymptotic behavior using Big-$O$ notation; here, $O(n^{-2})$ denotes a term bounded above by $n^{-2}$.
In addition, we define the residual term:
\begin{align} \label{eq:en;schmidt}
  \epsilon_n \define {a_n}^2 - \dfrac{1}{\pi(n+1/4)}
  = O(n^{-3}).
\end{align}
Substituting Eq.~\eqref{eq:en;schmidt} into Eq.~\eqref{eq:K-ratio_p;schmidt} yields the following asymptotic expression:
\begin{widetext}
\begin{align}
  A_p
  &= \dfrac{2}{\pi^2(p+1/2)} \sum_{n=0}^p
    \dfrac{1}{n+1/4}
  + \dfrac{2}{\pi (p+1/2)} \sum_{n=0}^p \epsilon_n \dfrac{p+1/2}{p-n+1/4}
  + \sum_{n=0}^{p} \epsilon_n \epsilon_{p-n}
    \nonumber \\ \label{eq:Ap-temp;schmidt}
  &= \dfrac{2}{\pi^2(p+1/2)}
    \mleft[ \psi\subt{d}\mleft( p + \dfrac{5}{4} \mright) - \psi\subt{d}\mleft( \dfrac{1}{4} \mright) \mright]
  + \dfrac{2}{\pi(p+1/2)}
    \mleft[
      \sum_{n=0}^p \epsilon_n
    + \sum_{n=0}^p \dfrac{n+1/4}{p-n+1/4} \epsilon_n
    \mright]
  + \sum_{n=0}^{p} \epsilon_n \epsilon_{p-n},
\end{align}
\end{widetext}
where $\psi\subt{d}(x)$ is the digamma function.
The two digamma functions appearing above are treated asymptotically or evaluated as follows~\cite{luke1969special,andrews1999special}:
\begin{align} \label{eq:dig_p+5/4;schmidt}
  \psi\subt{d}\mleft( p + \dfrac{5}{4} \mright)
  &= \ln\mleft( p + \dfrac{3}{4} \mright)
  + O(p^{-2}), \\ \label{eq:dig_1/4;schmidt}
  \psi\subt{d}\mleft( \dfrac{1}{4} \mright)
  &= - \gamma\subt{E} - \dfrac{\pi}{2} - 3\ln(2).
\end{align}
The sum involving $\epsilon_n$ appearing in Eq.~\eqref{eq:Ap-temp;schmidt} is evaluated by noting that for large $n$, $\epsilon_n=O(n^{-3})$. Thus, the truncation error introduced by replacing the finite sum with the corresponding infinite sum is $O(p^{-2})$, yielding:
\begin{align}
  \sum_{n=0}^p \epsilon_n
  &\approx \sum_{n=0}^\infty \epsilon_n
    \nonumber \\
  &= \lim_{x\to 1^-} \mleft[
      \sum_{n=0}^\infty {a_{n}}^2 x^n
      - \dfrac{1}{\pi} \sum_{n=0}^\infty \dfrac{x^n}{n+1/4}
    \mright] \nonumber \\ \label{eq:en-temp;schmidt}
  &= \lim_{x\to 1^-} \mleft(
      \dfrac{2}{\pi} K\subt{ell}(\sqrt{x})
      - \dfrac{4}{\pi}\pFq{2}{1}{\frac{1}{4},1}{\frac{5}{4}}{x}
    \mright),
\end{align}
where $K\subt{ell}$ is the complete elliptic integral of the first kind.
The two individual terms in Eq.~\eqref{eq:en-temp;schmidt} exhibit logarithmic divergences as $x\to 1^-$, but these divergences cancel out in their difference. Their respective limiting forms are
\begin{align}
  \lim_{x\to 1^-} \dfrac{2}{\pi} K\subt{ell}(\sqrt{x})
  &= -\dfrac{1}{\pi}\mleft[
    \lim_{x\to 1^-} \ln(1-x)
    - 4\ln(2) \mright], \nonumber \\
  \lim_{x\to 1^-} \dfrac{4}{\pi}
    \pFq{2}{1}{\frac{1}{4},1}{\frac{5}{4}}{x}
  &= - \dfrac{1}{\pi} \mleft[
    \lim_{x\to 1^-} \ln(1-x)
    - \dfrac{\pi}{2} - 3\ln(2)
    \mright].
\end{align}
Consequently, Eq.~\eqref{eq:en-temp;schmidt} evaluates to
\begin{align} \label{eq:en-sum;schmidt}
  \sum_{n=0}^p \epsilon_n
  &\approx \dfrac{\ln(2)}{\pi} - \dfrac{1}{2}.
\end{align}
Next, consider the correction term inside the bracket of the second contribution to $A_p$ in Eq.~\eqref{eq:Ap-temp;schmidt}. Since $\epsilon_n = O(n^{-3})$, the bound $(n+1/4)\epsilon_n = O(n^{-2})$ holds, ensuring the convergence of the infinite series $\sum_{n=0}^{\infty}(n+1/4)\epsilon_n$. It then follows that
\begin{align} \label{eq:second_corr;schmidt}
  \sum_{n=0}^p \dfrac{n+1/4}{p-n+1/4} \epsilon_n
  &= O(p^{-1}).
\end{align}
Combined with the prefactor $2/[\pi(p+1/2)]$, this correction contributes to $A_p$ only at order $O(p^{-2})$.

Finally, using $\epsilon_n=O(n^{-3})$, the convolution term in Eq.~\eqref{eq:Ap-temp;schmidt} satisfies
\begin{align} \label{eq:conv_err;schmidt}
  \sum_{n=0}^{p} \epsilon_n \epsilon_{p-n}
  &= O(p^{-3}).
\end{align}

Thus, both the correction term in Eq.~\eqref{eq:second_corr;schmidt} and Eq.~\eqref{eq:conv_err;schmidt} are subleading compared with the $O(p^{-1}\ln p)$ contribution arising from the digamma function in Eq.~\eqref{eq:dig_p+5/4;schmidt}. Collecting the leading contributions from Eqs.~\eqref{eq:dig_p+5/4;schmidt}, \eqref{eq:dig_1/4;schmidt}, and \eqref{eq:en-sum;schmidt} into Eq.~\eqref{eq:Ap-temp;schmidt}, the inverse Schmidt number ratio $A_p$ is approximated as
\begin{align}
  A_p
  &\approx \dfrac{2}{\pi^2 (p+1/2)} \mleft[
    \ln\mleft( p + \dfrac{3}{4} \mright)
    + \gamma\subt{E} + 4\ln(2)
    \mright],
\end{align}
leading to the final form:
\begin{align}
  \dfrac{K_{N_+}^{N_-}(\xi)}{K_0^0(\xi)}
  &\approx \dfrac{\pi^2}{2}
    \dfrac{p+1/2}{\ln(p+3/4)+\gamma\subt{E} + 4\ln(2)}.
\end{align}

\subsubsection{Generalization to high-order spatial modes with non-zero OAM}

We extend this derivation to the general case of spatial modes with non-zero OAM $l \neq 0$. We define the scale factor $A_p^l$ as the ratio of the fundamental-mode Schmidt number to the higher-order-mode Schmidt number:
\begin{align} \label{eq:K-ratio;schmidt}
  A_p^l
  &\define \dfrac{K_0^0(\xi)}{K_{N_+}^{N_-}(\xi)}
  \approx \sum_{n=0}^p {a_{n}}^2 a_{p-n} a_{p+\abs{l}-n}
\end{align}

To evaluate the summation in Eq.~\eqref{eq:K-ratio;schmidt} in the limit of large radial mode $p$, we define the continuous variable $x = n/(p+1/4)$ within the interval $0 \leq x < 1$ and introduce the ratio parameter $\alpha = \abs{l}/(p+1/4)$. This parametrization reformulates the summation as
\begin{align} \label{eq:K-ratio-appr;schmidt}
  A_p^l
  &= \dfrac{1}{\pi (p+1/4)} \sum_{n=0}^{p} {a_{n}}^2 h_\alpha (x)
  + O(p^{-3/2}),
\end{align}
where the weight function $h_\alpha(x)$ is defined as
\begin{align}
  h_\alpha (x)
  &= \dfrac{1}{\sqrt{(1-x)(1-x+\alpha)}}.
\end{align}
We partition Eq.~\eqref{eq:K-ratio-appr;schmidt} by subtracting and adding the boundary value $h_\alpha(0)$:
\begin{align} \label{eq:K-ratio-appr2;schmidt}
  A_p^l
  &= \dfrac{1}{\pi (p+1/4)\sqrt{1+\alpha}} \sum_{n=0}^{p}
    a_{n}^2
    \nonumber \\ & \quad
  + \dfrac{1}{\pi (p+1/4)} \sum_{n=1}^{p}
    a_{n}^2 [h_\alpha (x) - h_\alpha (0)]
  + O\mleft( \dfrac{\ln(p)}{p^2} \mright),
\end{align}
The first sum on the right-hand side of Eq.~\eqref{eq:K-ratio-appr2;schmidt} can be evaluated using Eqs.~\eqref{eq:en;schmidt} and \eqref{eq:Ap-temp;schmidt}, which lead to Eqs.~\eqref{eq:dig_1/4;schmidt} and \eqref{eq:en-sum;schmidt}, as follows
\begin{align} \label{eq:a2;schmidt}
  \sum_{n=0}^{p} {a_{n}}^2
  &= \dfrac{1}{\pi} \sum_{n=0}^p
    \mleft[ \dfrac{1}{n+1/4} + \epsilon_n \mright] \nonumber \\
  &\approx \dfrac{1}{\pi} \mleft[
      \ln\mleft( p + \dfrac{3}{4} \mright)
      + \gamma\subt{E} + 4\ln(2)
    \mright] + O(p^{-1}).
\end{align}
The second sum is approximated by a continuous integration:
\begin{align} \label{eq:bulk;schmidt}
  \sum_{n=1}^{p}
    a_{n}^2 [h_\alpha (x) - h_\alpha (0)]
  &\approx \dfrac{1}{\pi} \int_0^1 \dd{x}~
    \dfrac{h_\alpha (x) - h_\alpha (0)}{x}
    \nonumber \\
  &= \dfrac{1}{\pi\sqrt{1+\alpha}}
    \ln\mleft[ \dfrac{4(1+\alpha)}{\alpha} \mright]
\end{align}
Substituting Eqs.~\eqref{eq:a2;schmidt} and \eqref{eq:bulk;schmidt} into Eq.~\eqref{eq:K-ratio-appr2;schmidt} and converting the ratio parameter back to its physical representation $\abs{l}/(p+1/4)$ yields the final asymptotic expression for the scale factor:
\begin{align}
  \dfrac{K_0^0(\xi)}{K_{N_+}^{N_-}(\xi)}
  &\approx \dfrac{\ln\mleft[\frac{(p+3/4)(p+\abs{l}+1/4)}{\abs{l}}\mright] + \gamma\subt{E} + 6\ln(2)}{\pi^2 \sqrt{(p+1/4)(p+\abs{l}+1/4)}}.
\end{align}
Therefore, Schmidt number $K_{N_+}^{N_-}(\xi)$ is approximated by
\begin{align}
  \dfrac{K_{N_+}^{N_-}(\xi)}{K_0^0(\xi)}
  &= \dfrac{\pi^2 \sqrt{(p+1/4)(p+\abs{l}+1/4)}}{\ln\mleft[\frac{(p+3/4)(p+\abs{l}+1/4)}{\abs{l}}\mright] + \gamma\subt{E} + 6\ln(2)}.
\end{align}

\begin{acknowledgments}
  This work was supported by JSPS KAKENHI Grant Numbers 25K24777 and 26K22745.
  The authors gratefully acknowledge the fruitful discussions held during the 12th QUATUO Workshop, Japan, which significantly contributed to this work.
\end{acknowledgments}

\bibliography{references.bib}

\end{document}